\documentclass[nofootinbib,aps,11pt]{revtex4-1}
\usepackage{graphicx}
\usepackage{hyperref}
\usepackage{cancel}
\usepackage{amssymb}
\usepackage{textcomp}
\usepackage{amsmath}
\usepackage{bm}
\usepackage{times}
\usepackage{epsfig}
\usepackage{color}
\newcommand{\hc}{{\rm h.c.}}
\newcommand{\eV}{{\rm eV}}
\newcommand{\keV}{{\rm keV}}

\newcommand{\GeV}{{\rm GeV}}
\newcommand{\TeV}{{\rm TeV}}

\begin{document}
\title{\LARGE Leptogenesis and Dark Matter from Low Scale Seesaw }
\bigskip
\author{Ang Liu}
\author{Zhi-Long Han}
\email{sps\_hanzl@ujn.edu.cn}
\author{Yi Jin}
\author{Fa-Xin Yang}
\affiliation{
School of Physics and Technology, University of Jinan, Jinan, Shandong 250022, China}
\date{\today}

\begin{abstract}
In this paper, we perform a detail analysis on leptogenesis and dark matter form low scale seesaw. In the framework of $\nu$2HDM, we further introduce one scalar singlet $\phi$ and one Dirac fermion singlet $\chi$, which are charged under a $Z_2$ symmetry. Assuming the coupling of $\chi$ is extremely small, it serves as a FIMP dark matter. The heavy right hand neutrinos $N$ provide a common origin for tiny neutrino mass (via seesaw mechanism), leptogenesis (via $N\to \ell_L \Phi_\nu^*,\bar{\ell}_L \Phi_\nu$) and dark matter (via $N\to \chi\phi$). With hierarchical right hand neutrino masses, the explicit calculation shows that success thermal leptogenesis is viable even for TeV scale $N_1$ with $0.4~\GeV\lesssim v_\nu\lesssim1$ GeV and lightest neutrino mass $m_1\lesssim 10^{-11}$ eV.
In such scenario, light FIMP dark matter in the keV to MeV range is naturally expected. The common parameter space for neutrino mass, natural leptogenesis and FIMP DM is also obtained in this paper.
\end{abstract}

\maketitle

\section{Introduction}
Besides the success of standard model (SM), there are still several open questions. In particular, tiny neutrino mass, baryon asymmetry of the Universe (BAU) and dark matter (DM) are the three outstanding evidences that require physics beyond SM. The discovery of neutrino oscillations  \cite{Fukuda:1998mi,Ahmad:2002jz} indicate that the mass of neutrinos are at sub-eV scale, which is at least six order of magnitudes smaller than charged leptons. Known as type-I seesaw mechanism \cite{Minkowski:1977sc,Mohapatra:1979ia}, this extensively considered way to naturally incorporate neutrino masses is via introducing three right hand  neutrinos $N$ together with high scale Majorana masses of $N$,
\begin{equation}
-\mathcal{L}_Y\supset y\overline{L} \widetilde{\Phi} N   +\frac{1}{2}\overline{N ^c}m_{N } N + \hc,
\end{equation}
where $\Phi$ is the SM Higgs doublet. After spontaneous electroweak symmetry breaking, neutrinos achieve masses as
\begin{equation}
m_\nu = - \frac{v^2}{2} y~ m_{N}^{-1} y^T.
\end{equation}
Typically, $m_\nu\sim\mathcal{O}(0.1)$ eV is obtained with $y\sim\mathcal{O}(1)$ and $m_N\sim\mathcal{O}(10^{14})$ GeV. Meanwhile, the heavy neutrino can also account for BAU via leptogenesis \cite{Fukugita:1986hr}. For canonical thermal leptogenesis with hierarchal right hand neutrinos, an upper limit on the CP asymmetry exists, thus a lower limit on mass of lightest right hand neutrino $M_1$ should be satisfied \cite{Davidson:2002qv},
\begin{equation}
M_1\gtrsim 5\times10^8 ~\GeV \left(\frac{v}{246~\GeV}\right)^2.
\end{equation}
Therefore, both tiny neutrino mass and leptogenesis favor high scale $N$ in type-I seesaw. However for such high scale $N$, a naturalness problem might arise \cite{Vissani:1997ys}. By requiring radiative corrections to the $m_\Phi^2 \Phi^\dag\Phi$ term no larger than $1~\TeV^2$, it is found that\cite{Clarke:2015gwa}
\begin{equation}\label{eq:M1}
M_1\lesssim3\times10^7 ~\GeV \left(\frac{v}{246~\GeV}\right)^{2/3},
\end{equation}
should be satisfied. Clearly, naturalness is incompatible with leptogenesis.
One viable pathway to overcome this is lowering the leptogenesis scale by imposing resonant leptogenesis \cite{Pilaftsis:2003gt}, ARS mechanism via neutrino oscillation \cite{Akhmedov:1998qx,Asaka:2005pn}, or from Higgs decays \cite{Hambye:2016sby,Hambye:2017elz}. All the success of these scenarios depend on the degenerate mass of right hand neutrinos \cite{Baumholzer:2018sfb}, which seems is another sense of unnatural. An alternative scenario with hierarchal right hand neutrinos is employing intrinsic low scale neutrino mass model, e.g., $\nu$2HDM \cite{Chao:2012pt,Clarke:2015hta} or Scotogenic model \cite{Ma:2006fn,Kashiwase:2012xd,Kashiwase:2013uy,Racker:2013lua,
Hugle:2018qbw,Borah:2018uci,Borah:2018rca,Mahanta:2019sfo}. In this paper, we consider the $\nu$2HDM \cite{Ma:2000cc}. Based on previous brief discussion in Ref.~\cite{Haba:2011ra,Haba:2011yc,Clarke:2015hta}, we perform a detailed analysis on leptogenesis, especially focus on dealing with the corresponding Boltzmann equations to obtain the viable parameter space.

On the other hand, dark matter accounts for more than five times the proportion of visible baryonic matter in our current cosmic material field. In principle, one can regard the lightest right hand neutrino $N_1$ at keV scale as sterile neutrino DM \cite{Dodelson:1993je,Adhikari:2016bei,Adulpravitchai:2015mna,Han:2018pek}. However, various constraints leave a quite small viable parameter space \cite{Boyarsky:2018tvu}. Meanwhile, leptogenesis with two hierarchal right hand neutrinos is actually still at high scale \cite{Hugle:2018qbw,Antusch:2011nz,Mahanta:2019gfe}.
In this paper, we further introduce a dark sector with one scalar singlet $\phi$ and one Dirac fermion singlet $\chi$, which are charged under a $Z_2$ symmetry \cite{Chianese:2019epo}. The stability of DM $\chi$  is protected by the $Z_2$ symmetry, therefore the tight X-ray limits can be avoided \cite{Boyarsky:2018tvu}. In light of the null results from DM direct detection \cite{Aprile:2018dbl} and indirect detection \cite{Ackermann:2015zua}, we consider $\chi$ as a FIMP DM \cite{Bernal:2017kxu}.

The structure of the paper is as follows. In Sec. \ref{SEC:TM}, we briefly introduce our model. Leptogenesis with hierarchal right hand neutrinos is discussed in Sec. \ref{SEC:LG}. The relic abundance of FIMP DM $\chi$ and constraint from free streaming length are considered in Sec. \ref{SEC:DM}.
Viable parameter space for leptogenesis and DM is obtained by a random scan in Sec. \ref{SEC:CA}. We conclude our work in Sec. \ref{SEC:CL}.

\section{The Model}\label{SEC:TM}

The original TeV-scale $\nu$2HDM for neutrino mass was proposed in Ref.~\cite{Ma:2000cc}. The model is extended by one neutrinophilic scalar doublet $\Phi_\nu$ with same quantum numbers as SM Higgs doublet $\Phi$ and three right hand heavy neutrino $N$. To forbid the direct type-I seesaw interaction $\overline{L}\tilde{\Phi}N$, a global $U(1)_L$ symmetry should be employed, under which $L_\Phi=0$, $L_{\Phi_\nu}=-1$ and $L_{N}=0$. Therefore,  $\Phi_\nu$ will specifically couple to $N$, and $\Phi$ couple to quarks and charge leptons as in SM. For the dark sector, one scalar singlet $\phi$ and one Dirac fermion singlet $\chi$ are further introduced, which are charged under a  $Z_2$ symmetry. Provided $m_\chi<m_\phi$, then $\chi$ serves as DM candidate.

The scalar doublets could be denoted as
\begin{align}
\Phi=\left(
\begin{array}{c}
\phi^+\\
\frac{v+\phi^{0,r}+i \phi^{0,i}}{\sqrt{2}}
\end{array}\right),~
\Phi_\nu=\left(
\begin{array}{c}
\phi^+_\nu\\
\frac{v_{\nu}+\phi^{0,r}_{\nu}+i \phi^{0,i}_{\nu}}{\sqrt{2}}
\end{array}\right).
\end{align}
The corresponding Higgs potential is then
\begin{eqnarray}
V & = & m_{\Phi}^2 \Phi^\dag \Phi +  m_{\Phi_\nu}^2 \Phi^\dag_\nu \Phi_\nu
       +m_\phi^2 \phi^\dag\phi+\frac{\lambda_1}{2} (\Phi^\dag \Phi)^2+\frac{\lambda_2}{2} (\Phi^\dag_\nu \Phi_\nu)^2\\ \nonumber
 & & +\lambda_3 (\Phi^\dag \Phi)(\Phi^\dag_\nu \Phi_\nu)+\lambda_4(\Phi^\dag \Phi_\nu)(\Phi^\dag_\nu \Phi)   - (\mu^2 \Phi^\dag\Phi_\nu +\hc )\\
 && +\frac{\lambda_5}{2} (\phi^\dag\phi)^2 + \lambda_6 (\phi^\dag \phi)(\Phi^\dag\Phi)+\lambda_7 (\phi^\dag \phi) (\Phi_\nu^\dag\Phi_\nu),
\end{eqnarray}
where the $U(1)_L$ symmetry is broken explicitly but softly by the $\mu^2$ term.
For the unbroken $Z_2$ symmetry, $\langle\phi\rangle=0$ should be satisfied. Meanwhile, VEVs of Higgs doublets in terms of parameters of the Higgs potential can be found by deriving the minimization condition
\begin{eqnarray}
v\left[m_{\Phi}^2+\frac{\lambda_1}{2} v^2 +\frac{\lambda_3+\lambda_4}{2}v_\nu^2\right]-\mu^2 v_\nu=0 \\
v_\nu\left[m_{\Phi_\nu}^2+\frac{\lambda_2}{2} v_\nu^2 +\frac{\lambda_3+\lambda_4}{2}v^2\right]-\mu^2 v=0.
\end{eqnarray}
Taking the parameter set
\begin{equation}
m_{\Phi}^2<0, m_{\Phi_\nu}^2>0, |\mu^2|\ll m_{\Phi_\nu}^2,
\end{equation}
we can obtain the relations of VEVs as
\begin{equation}
v\simeq \sqrt{\frac{-2 m_{\Phi}^2}{\lambda_1}}, v_\nu \simeq \frac{\mu^2 v}{m_{\Phi_\nu}^2+(\lambda_3+\lambda_4)v^2/2}.
\end{equation}
Typically, $v_\nu\sim1$ GeV is obtained with $\mu\sim10~\GeV$ and $m_{\Phi_\nu}\sim100$ GeV. Since $\mu^2$ term is the only source of $U(1)_L$ breaking, radiative corrections to $\mu^2$ are proportional to $\mu^2$ itself and are only logarithmically sensitive to the cutoff \cite{Davidson:2009ha}. Thus, the VEV hierarchy $v_\nu\ll v$ is stable against radiative corrections \cite{Morozumi:2011zu,Haba:2011fn}.

After SSB, the physical Higgs bosons are given by \cite{Guo:2017ybk}
\begin{eqnarray}
H^+=\phi^+_\nu\cos\beta-\phi^+\sin\beta&,~& A=\phi^{0,i}_\nu\cos\beta-\phi^{0,i}\sin\beta, \\
H=\phi^{0,r}_\nu\cos\alpha-\phi^{0,r}\sin\alpha&,~& h=\phi^{0,r}\cos\alpha+\phi^{0,r}\sin\alpha,
\end{eqnarray}
where the mixing angles $\beta$ and $\alpha$  are determined by
\begin{equation}\label{mix}
\tan\beta=\frac{v_\nu}{v},~\tan2\alpha\simeq2\frac{v_\nu}{v}
\frac{-\mu^2+(\lambda_3+\lambda_4)vv_\nu}{-\mu^2+\lambda_1 vv_\nu}.
\end{equation}
Neglecting terms of  $\mathcal{O}(v_\nu^2)$ and $\mathcal{O}(\mu^2)$, masses of the physical Higgs bosons are
\begin{eqnarray}
m_{H^+}^2\simeq m_{\Phi_\nu}^2\!+\frac{1}{2}\lambda_3v^2,~m_A^2\simeq m_H^2\simeq m_{H^+}^2\!+\frac{1}{2}\lambda_4v^2,~ m_h^2 \simeq  \lambda_1 v^2.
\end{eqnarray}
Since the mixing angles are suppressed by the small value of $v_\nu$, $h$ is almost identically to the $125$ GeV SM Higgs boson \cite{Aad:2012tfa,Chatrchyan:2012xdj}. A degenerate mass spectrum of $\Phi_\nu$ as $m_{H^+}\!=\!m_{H}\!=\!m_A\!=\!m_{\Phi_\nu}$ is adopted in our following discussion for simplicity, which is certainly allowed by various constraints ~\cite{Machado:2015sha}.  Due to the unbroken $Z_2$ symmetry, the dark scalar singlet $\phi$ do not mix with the Higgs doublets.

The new Yukawa interaction and  mass terms are
\begin{equation}\label{yuk}
-\mathcal{L}_Y\supset y\overline{L} \widetilde{\Phi}_\nu N  +\lambda \bar{\chi}\phi N  +\frac{1}{2}\overline{N ^c}m_{N } N  + m_\chi \bar{\chi} \chi+ \hc,
\end{equation}
where $\widetilde{\Phi}_\nu=i\sigma_2 \Phi_\nu^*$.
Similar to the canonical Type-I seesaw \cite{Minkowski:1977sc},  the mass matrix for light neutrinos can be derived from Eq. (\ref{yuk}) as:
\begin{equation}\label{eq:mv}
m_\nu = - \frac{v_\nu^2}{2} y~ m_{N}^{-1} y^T = U_{\text{PMNS}}\, \hat{m}_\nu U^T_{\text{PMNS}},
\end{equation}
where $\hat{m}_\nu=\mbox{diag}(m_1,m_2,m_3)$ is the diagonalized neutrino mass matrix, and $U_{\text{PMNS}}$ is the PMNS (Pontecorvo-Maki-Nakagawa-Sakata) matrix:
\begin{align}
U_{\text{PMNS}}\! =\! \left(
\begin{array}{ccc}
c_{12} c_{13} & s_{12} c_{13} & s_{13} e^{i\delta}\\
-s_{12}c_{23}-c_{12}s_{23}s_{13}e^{-i\delta} & c_{12}c_{23}-s_{12}s_{23}s_{13} e^{-i\delta} & s_{23}c_{13}\\
s_{12}s_{23}-c_{12}c_{23}s_{13}e^{-i\delta} & -c_{12}s_{23}-s_{12}c_{23}s_{13}e^{-i\delta} & c_{23}c_{13}
\end{array}
\right)\!\times\!
\text{diag}(e^{i \varphi_1/2},1,e^{i\varphi_2/2})
\end{align}
Here, we use abbreviations $c_{ij}=\cos\theta_{ij}$ and $s_{ij}=\sin\theta_{ij}$, $\delta$ is the Dirac phase and $\varphi_1,\varphi_2$ are the two Majorana phases. Due to smallness of $v_\nu$, TeV scale $m_{N}$ could be viable to realise $0.1$ eV scale light neutrino masses. Using the Casas-Ibarra parametrization \cite{Casas:2001sr,Ibarra:2003up}, the Yukawa matrix $y$ can be expressed in terms of neutrino oscillation parameters
\begin{equation}\label{eq:CI}
y=\frac{\sqrt{2}}{v_\nu}U_{\text{PMNS}}\hat{m}_\nu^{1/2} R (\hat{m}_{N})^{1/2},
\end{equation}
where $R$ is an orthogonal matrix in general and $\hat{m}_N=\mbox{diag}(M_1,M_2,M_3)$ is the diagonalized heavy neutrino mass matrix. In this work, we parameterize matrix $R$ as
\begin{align}
R = \left(
\begin{array}{ccc}
\cos\omega_{12} & -\sin\omega_{12} & 0 \\
\sin\omega_{12} & \cos\omega_{12} & 0 \\
0 & 0& 1
\end{array}
\right)
\left(
\begin{array}{ccc}
\cos\omega_{13} & 0& -\sin\omega_{13} \\
0 & 1 & 0 \\
\sin\omega_{13} & 0& \cos\omega_{13}
\end{array}
\right)
\left(
\begin{array}{ccc}
1 & 0 & 0 \\
0 & \cos\omega_{23} & -\sin\omega_{23} \\
0 & \sin\omega_{23}& \cos\omega_{23}
\end{array}
\right),
\end{align}
where $\omega_{12,13,23}$ are arbitrary complex angles.

\section{Leptogenesis}\label{SEC:LG}

Now we consider the leptogenesis in this model. The lepton asymmetry is generated by the out-of-equilibrium CP-violating decays of right hand neutrino $N\to \ell_L \Phi_\nu^*,\bar{\ell}_L \Phi_\nu$. Neglecting the flavor effect \cite{Nardi:2006fx}, the CP asymmetry is given by
\begin{equation}
\epsilon_i=\frac{1}{8\pi (y^\dag y)_{ii}} \sum_{j\neq i} \text{Im}[(y^\dag y)^2_{ij}] G\left(\frac{M_j^2}{M_i^2},\frac{m_{\Phi_\nu}^2}{M_i^2}\right),
\end{equation}
where the function $G(x,y)$ is defined as \cite{Mahanta:2019gfe}
\begin{equation}
G(x,y)=\sqrt{x}\left[\frac{(1-y)^2}{1-x}+1+\frac{1-2y+x}{(1-y^2)^2} \ln
\left(\frac{x-y^2}{1-2y+x}\right)\right].
\end{equation}
Using the parametrization of Yukawa coupling $y$ in Eq.~(\ref{eq:CI}), it is easy to verify
\begin{equation}\label{eq:yy}
y^\dag y = \frac{2}{v_\nu^2} \hat{m}_{N}^{1/2} R^\dag \hat{m}_\nu R
\hat{m}_{N}^{1/2}.
\end{equation}
Hence, the matrix $y^\dag y$ does not depend on the PMNS  matrix, which means that the complex matrix $R$ is actually the source of CP asymmetry $\epsilon_i$.
The asymmetry is dominantly generated by the decay of $N_1$. Further considering the hierarchal mass spectrum $m_{\Phi_\nu}^2\ll M_1^2\ll M_{2,3}^2$, the asymmetry $\epsilon_1$ is simplified to
\begin{equation}
\epsilon_1\simeq -\frac{3}{16\pi(y^\dag y)_{11}}\sum_{j=2,3} \text{Im}[(y^\dag y)^2_{1j}] \frac{M_1}{M_j}
\end{equation}
Similar to the Davidson-Ibarra bound \cite{Davidson:2002qv}, an upper limit on $\epsilon_1$ can be derived
\begin{equation}\label{eq:eps1}
|\epsilon_1|\lesssim \frac{3}{16\pi}\frac{M_1 m_3}{v_\nu^2}.
\end{equation}
Comparing with the bound in type-I seesaw, the asymmetry could be enhanced due to the smallness of VEV $v_\nu$. Therefore, low scale leptogenesis seems to be viable in the $\nu$2HDM \cite{Haba:2011ra,Clarke:2015hta}. Meanwhile, the washout effect is quantified by the decay parameter
\begin{equation}
K = \frac{\Gamma_1}{H(z=1)},
\end{equation}
where $\Gamma_1$ is the decay width of $N_1$, $H$ is the Hubble parameter and $z\equiv M_1/T$ with $T$ being the temperature of the thermal bath. The decay width is given by
\begin{equation}
\Gamma_1=\frac{M_1}{8\pi}(y^\dag y)_{11}\left(1-\frac{m_{\Phi_\nu}^2}{M_1^2}\right)^2,
\end{equation}
and the Hubble parameter is
\begin{equation}
H=\sqrt{\frac{8\pi^3 g_*}{90}}\frac{T^2}{M_{pl}}=H(z=1)\frac{1}{z^2},
\end{equation}
with $g_*$ the effective number of relativistic degrees of freedom and $M_{pl}=1.2\times10^{19}$ GeV. Using Eq. (\ref{eq:yy}), one can verify
\begin{equation}
K\simeq 897 \left(\frac{v}{v_\nu}\right)^2
\frac{(\hat{m}_\nu^R)_{11}}{\eV},
\end{equation}
where $\hat{m}_\nu^R\equiv R^\dag \hat{m}_\nu R$, and thus
\begin{equation}\label{eq:mvR}
(\hat{m}_\nu^R)_{11}=m_1 |\cos\omega_{12}|^2|\cos\omega_{13}|^2+m_2 |\sin\omega_{12}|^2|\cos\omega_{13}|^2+m_3|\sin\omega_{13}|^2.
\end{equation}
It is obvious that the decay parameter $K$ does not depend on $\omega_{23}$, and it is also enhanced by smallness of $v_\nu$.
Since $(\hat{m}_\nu^R)_{11}$ is typically of the order of $m_3\sim0.1$ eV, the
decay parameter $K\simeq 5.4\times10^6$ when $v_\nu=1$ GeV. So even with maximum asymmetry $\epsilon_1^\text{max}\sim-6.0\times10^{-7}$ for $M_1=10^5$ GeV obtained from Eq.~\eqref{eq:eps1}, a rough estimation of final baryon asymmetry gives $Y_{\Delta B}\sim-10^{-3} \epsilon_1^\text{max}/K\sim1.1\times10^{-16}$ for strong washout \cite{Davidson:2008bu}, which is far below current observed value $Y_{\Delta B}^\text{obs}=(8.72\pm0.04)\times10^{-11}$ \cite{Aghanim:2018eyx}. Hence, only obtaining an enhanced CP asymmetry $\epsilon_1$ is not enough, one has to deal with the washout effect more carefully.

\begin{figure}
\begin{center}
\includegraphics[width=0.45\linewidth]{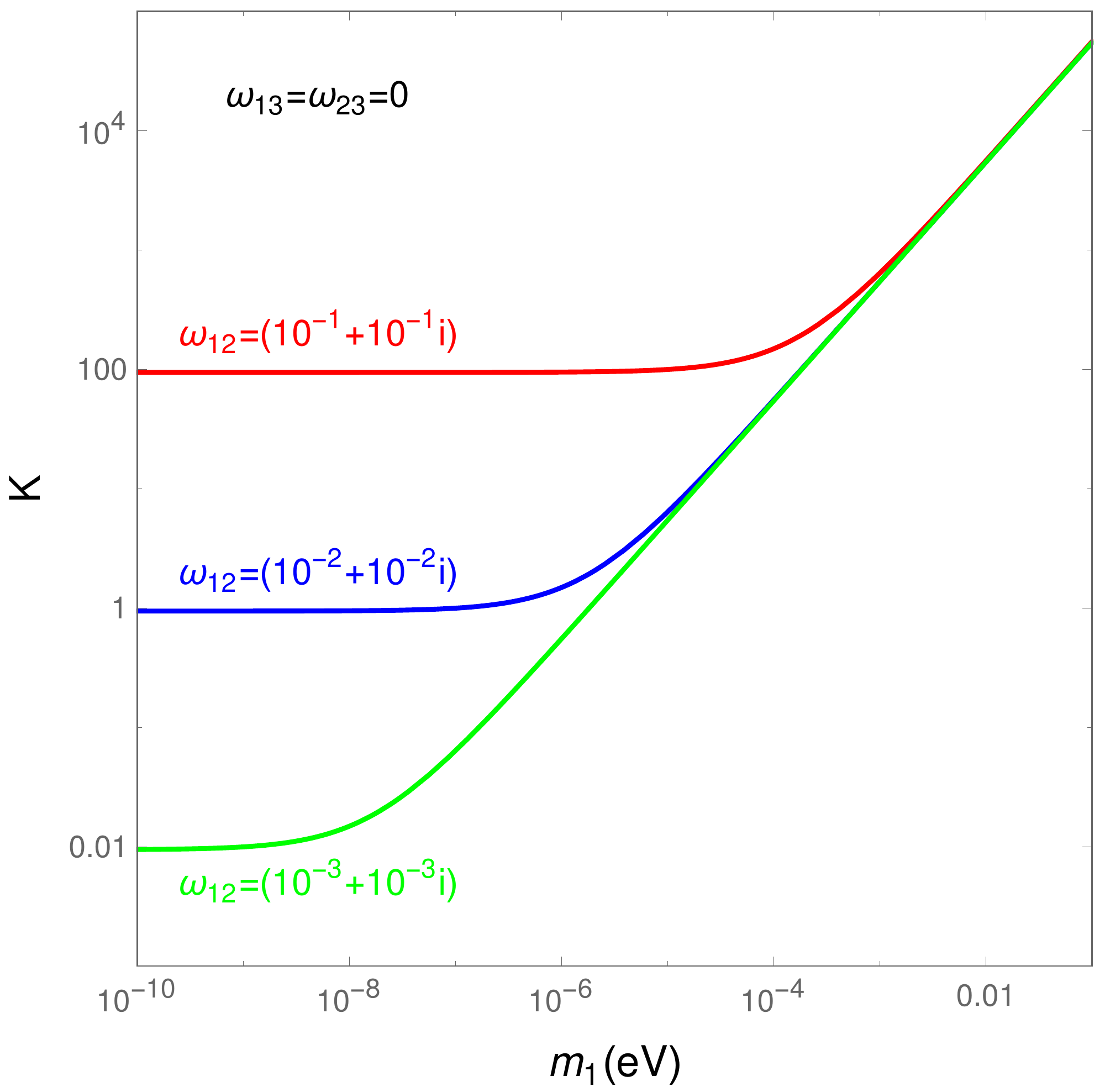}
\includegraphics[width=0.45\linewidth]{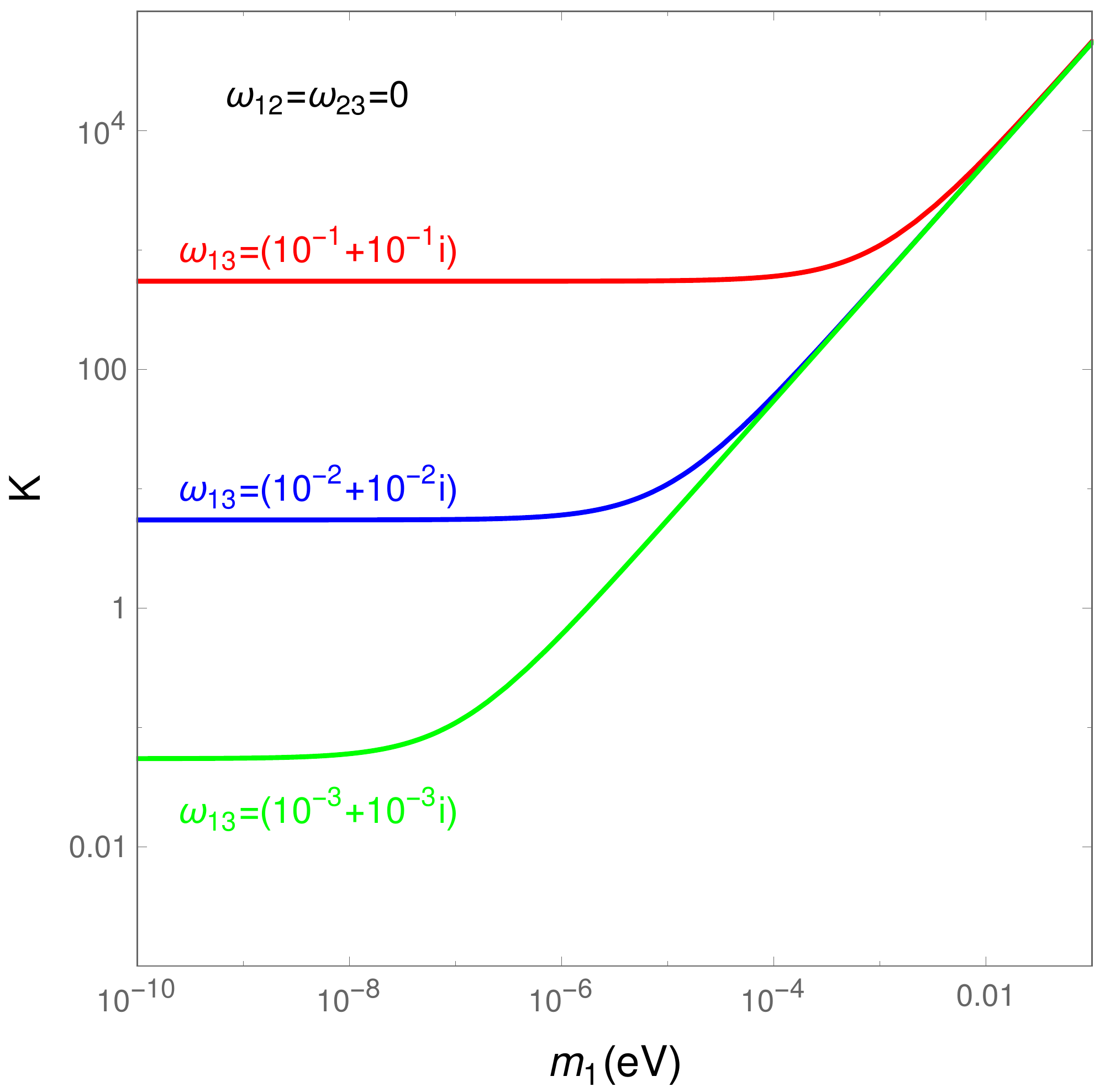}
\end{center}
\caption{Decay parameter $K$ as a function of $m_1$ with $v_\nu=10$ GeV. Because $R$ must be a complex matrix, we have set $\omega_{ij R}=\omega_{ij I}$.}
\label{FIG:K-m1}
\end{figure}

One promising pathway is to reduce the decay parameter $K$. For instance, if weak washout condition $K\lesssim1$ is realised, then $Y_{\Delta B}\sim-10^{-3} \epsilon_1^\text{max}\sim 6.0\times10^{-10}>Y_{\Delta B}^\text{obs}$. Thus, correct baryon asymmetry can be obtained by slightly tunning  $\epsilon_1$. As pointed out in Ref.~\cite{Hugle:2018qbw}, small value of $K$ can be realised by choosing small $\omega_{12,13}$. In Fig.~\ref{FIG:K-m1}, we illustrate the  dependence of $K$ on lightest neutrino mass $m_1$ with $v_\nu=10$ GeV. The left panel shows the special case $\omega_{13}=0$, where Eq.~\eqref{eq:mvR} is simplified to $(\hat{m}_\nu^R)_{11}=m_1 |\cos\omega_{12}|^2+m_2|\sin\omega_{12}|^2\geq \sqrt{\Delta m_{21}^2} |\sin\omega_{12}|^2$. It is clear that the weak washout condition $K<1$ favors $|\omega_{12}|\lesssim10^{-2}$ and $m_1\lesssim10^{-6}$ eV. The right panel shows the special case $\omega_{12}=0$. Similar results are observed with left panel.

On the other hand, the $\Delta L=2$ washout processes become more significant for small $v_\nu$ \cite{Haba:2011ra,Clarke:2015hta}. Notably, for low scale seesaw, the narrow width condition $\Gamma_1/M_1\ll1$ is satisfied. Therefore, the  evolution of lepton asymmetry and DM abundance actually decouple from each other \cite{Falkowski:2011xh,Falkowski:2017uya}. The evolution of abundance $Y_{N_1}$ and lepton asymmetry $Y_{\Delta L}$ are described by the Boltzmann equations
\begin{eqnarray}
\frac{dY_{N_1}}{dz} &=& - D (Y_{N_1}-Y_{N_1}^{eq}),\\
\frac{dY_{\Delta L}}{dz} &=&- \epsilon_1 D (Y_{N_1}-Y_{N_1}^{eq})
- W Y_{\Delta L}.
\end{eqnarray}
The decay term is given by
\begin{equation}
D = K z \frac{\mathcal{K}_1(z)}{\mathcal{K}_2(z)}.
\end{equation}
For the washout term, two contributions are considered, i.e., $W=W_{ID}+W_{\Delta L=2}$, where the inverse decay term is
\begin{equation}
W_{ID}= \frac{1}{4} K z^3 \mathcal{K}_1(z),
\end{equation}
and the $\Delta L=2$ scattering term at low temperature is approximately \cite{Buchmuller:2004nz}
\begin{equation}
W_{\Delta L=2}\simeq \frac{0.186}{z^2} \left(\frac{246~\GeV}{v_\nu}\right)^4
\left(\frac{M_1}{10^{10}~\GeV}\right) \left(\frac{\bar{m}}{\eV}\right)^2.
\end{equation}
Here, $\bar{m}$ is the absolute neutrino mass scale, which is calculated as
\begin{equation}
\bar{m}^2=m_1^2+m_2^2+m_3^2 =3m_1^2+ \Delta m^2_{21} + \delta m^2_{31},
\end{equation}
for normal hierarchy. According to latest global fit, we use the best fit values, i.e., $\Delta m^2_{21}=7.39\times10^{-5}~\eV^2$ and $\delta m^2_{31}=2.525\times10^{-3}~\eV^2$ \cite{Esteban:2018azc}. For tiny lightest neutrino mass $m_1\ll10^{-2}$ eV, we actually have $\bar{m}\simeq\sqrt{\delta m_{31}^2}\sim0.05$ eV. Notably, the $\Delta L=2$ scattering term would be greatly enhanced when $v_\nu\ll v$, so this term is much more important than in vanilla leptogenesis.
Then, the sphaleron processes convert the lepton asymmetry into baryon asymmetry as \cite{Harvey:1990qw}
\begin{equation}
Y_{\Delta B}= \frac{28}{79} Y_{\Delta(B-L)}=-\frac{28}{51}Y_{\Delta L}.
\end{equation}

\begin{figure}
\begin{center}
\includegraphics[width=0.45\linewidth]{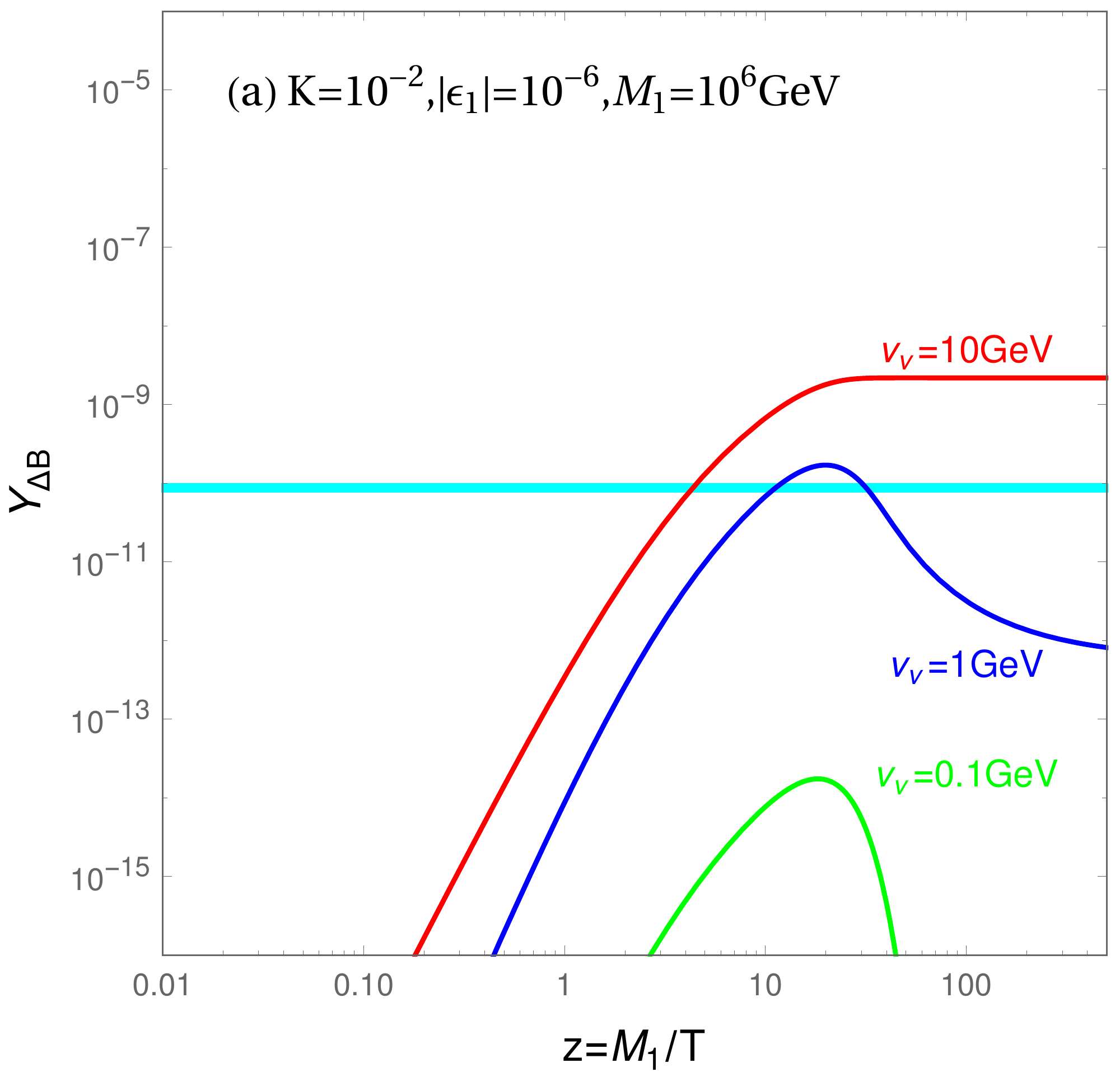}
\includegraphics[width=0.45\linewidth]{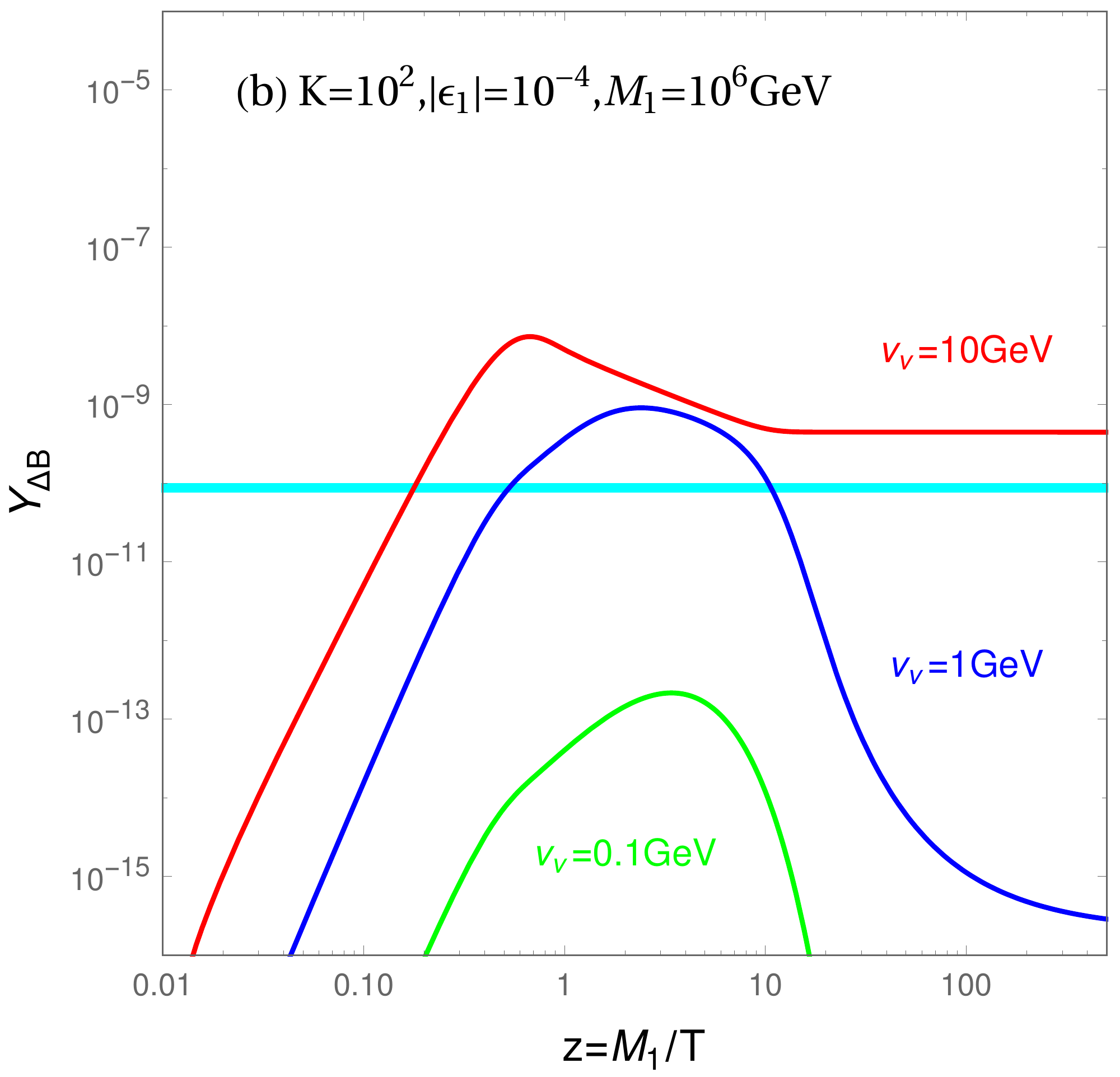}
\\
\includegraphics[width=0.45\linewidth]{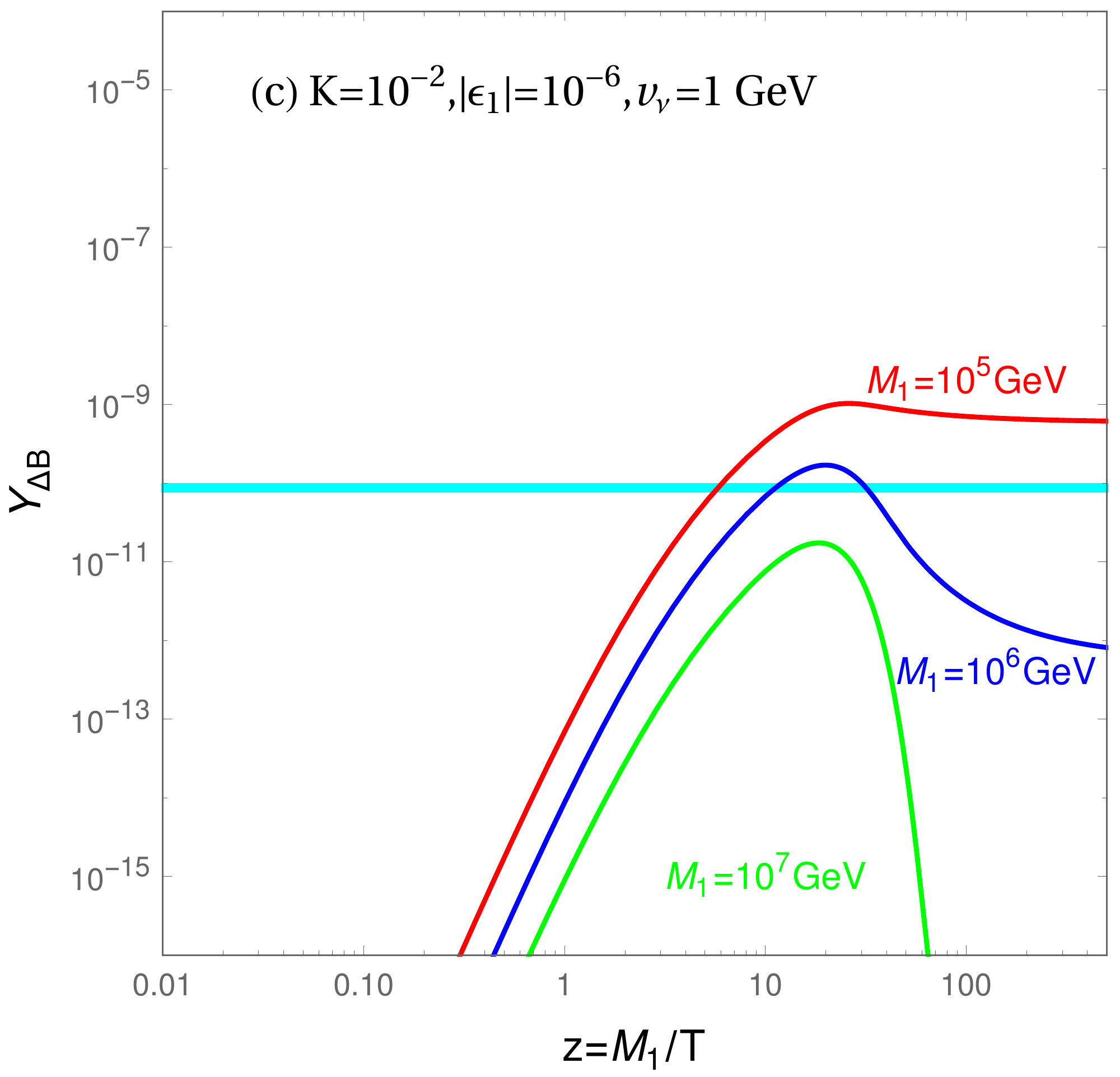}
\includegraphics[width=0.45\linewidth]{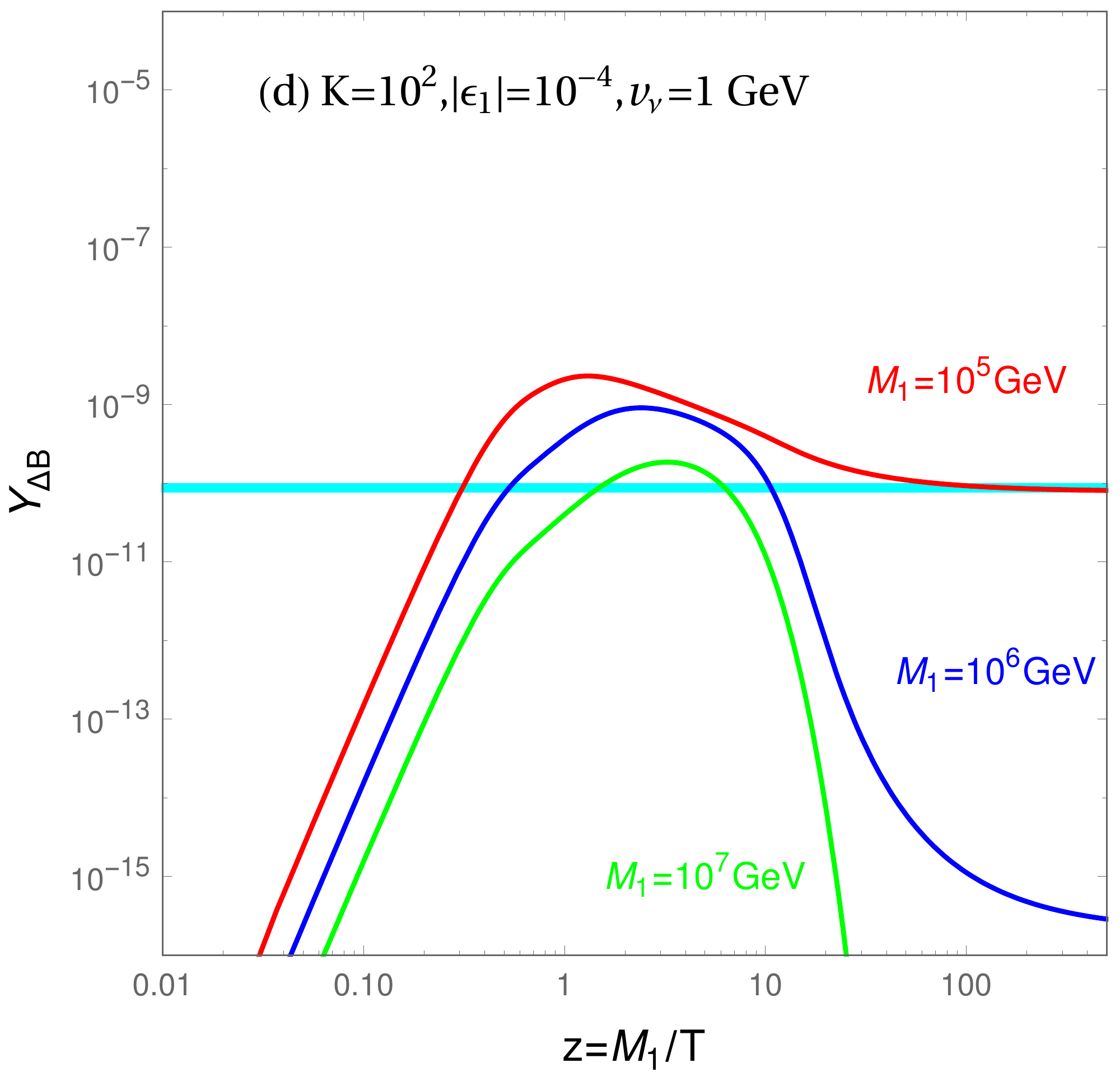}
\end{center}
\caption{The washout effect of $\Delta L=2$ processes. The cyan lines are the observed value $Y_{\Delta B}^\text{obs}=8.72\times10^{-11}$.}
\label{FIG:YB-Z}
\end{figure}

Fig.~\ref{FIG:YB-Z} shows the washout effect of $\Delta L=2$ processes. In Fig.~\ref{FIG:YB-Z} (a), weak washout scenario is considered by fixing $K=10^{-2},|\epsilon_1|=10^{-6}, M_1=10^6~\GeV$ while varying $v_\nu=10,1,0.1~\GeV$. It shows that for $v_\nu=10~\GeV$, the $\Delta L=2$ effect is not obvious, but for $v_\nu=1~\GeV$, the final baryon asymmetry $Y_{\Delta B}$ is diluted by over three orders of magnitude. While for $v_\nu=0.1~\GeV$, the $\Delta L=2$ effect is so strong that the final baryon asymmetry is negligible.
The strong washout scenario with $K=10^2, |\epsilon_1|=10^{-4}, M_1=10^6~\GeV$ and varying $v_\nu=10,1,0.1~\GeV$ is illustrated in Fig.~\ref{FIG:YB-Z} (b), where the final baryon asymmetry $Y_{\Delta B}$ for $v_\nu=1~\GeV$ is decreased by about six orders comparing with the case for $v_\nu=10~\GeV$. Therefore, the $\Delta L=2$ washout effects set a lower bound on $v_\nu$, i.e., $v_\nu\gtrsim0.3~\GeV$ as suggested by Ref.~\cite{Clarke:2015hta}. Furthermore, since the $\Delta L=2$ washout term is also proportional to $M_1$, the larger $M_1$ is, the more obvious the washout effect is. The corresponding results are depicted in Fig.~\ref{FIG:YB-Z} (c) for the weak washout and Fig.~\ref{FIG:YB-Z}~(d) for the strong washout. In this way, for certain value of $v_\nu$, an upper bound on $M_1$ can be obtained. For instance, when $v_\nu=1~\GeV$, then $M_1\lesssim10^5~\GeV$ should be satisfied \cite{Haba:2011ra}.

\section{Dark Matter}\label{SEC:DM}

In our extension of the $\nu$2HDM, the right-handed heavy neutrinos $N$ also couple with fermion singlet $\chi$ and scalar singlet $\phi$ via the Yukawa interaction. The complex Yukawa coupling coefficient $\lambda$ can lead to CP violation in $N$ decays, and eventually producing asymmetric DM $\chi$ \cite{Falkowski:2011xh}. Instead, we consider another interesting scenario, i.e., the FIMP case with the real coupling $\lambda \ll1$ \cite{Falkowski:2017uya}. In this way, the interaction of DM $\chi$ is so weak that it never reach thermalization. Its relic abundance is determined by the freeze-in mechanism \cite{Hall:2009bx}, which is obtained by solving the following Boltzmann equation
\begin{eqnarray}\label{eq:BE}
\frac{dY_{\chi}}{dz} &=&  D ~Y_{N_1} \text{BR}_\chi,
\end{eqnarray}
where $\text{BR}_\chi$ is the branching ratio of $N_1\to \chi \phi$. Due to the FIMP nature of $\chi$, the hierarchal condition $\text{BR}_\chi\ll\text{BR}_\ell\simeq1$ is easily satisfied.
The out of equilibrium condition for $N_1\to\chi\phi$ decay is $\Gamma_\chi/H(z=1)\simeq \text{BR}_\chi \Gamma_1/H(z=1)=\text{BR}_\chi K<1$.
In following studies, we mainly take $\text{BR}_\chi<10^{-2}$ and $K\lesssim10$, thus the out of equilibrium condition is always satisfied.
\begin{figure}
\centering
\includegraphics[width=0.45\textwidth]{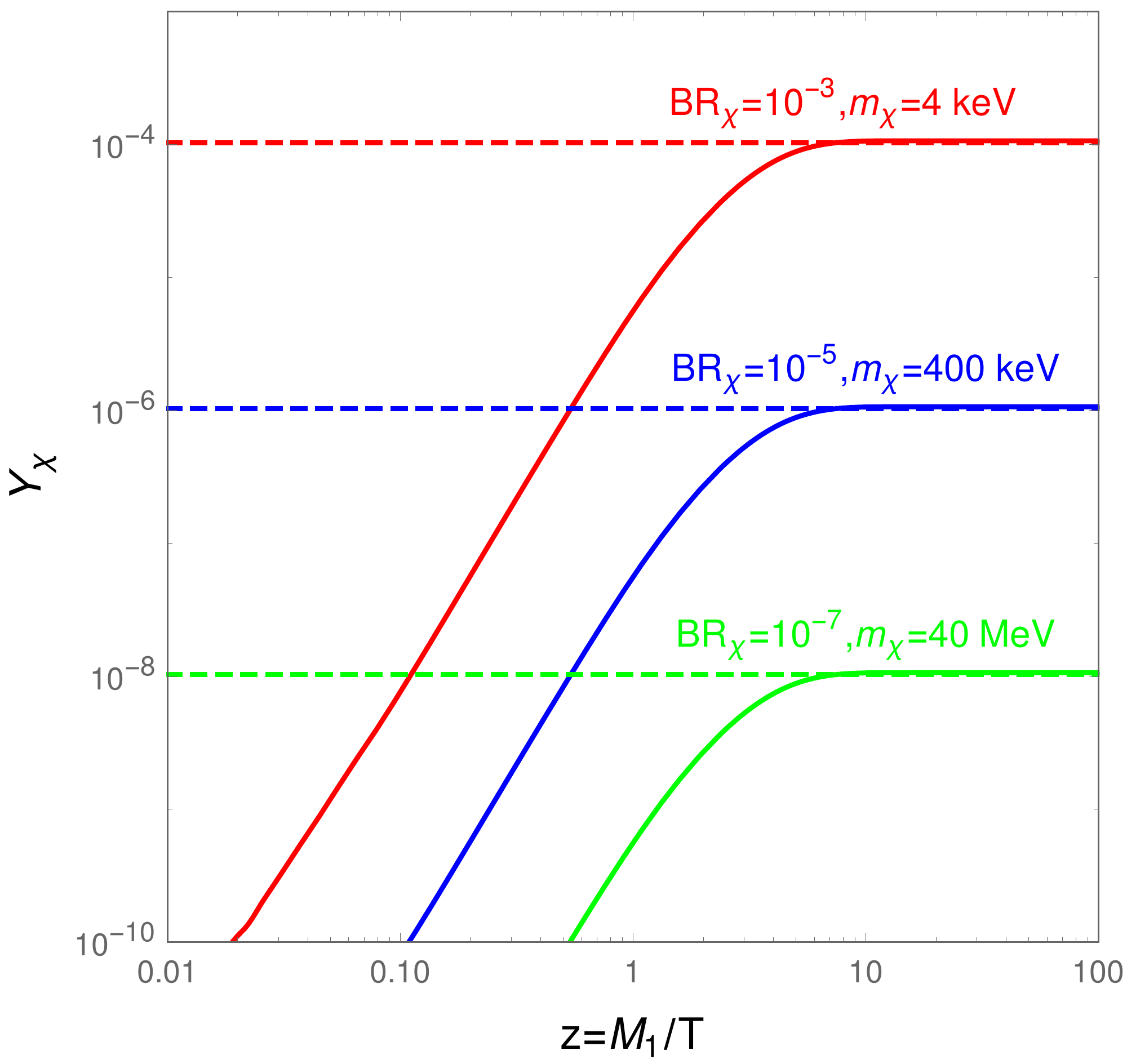}
\includegraphics[width=0.45\textwidth]{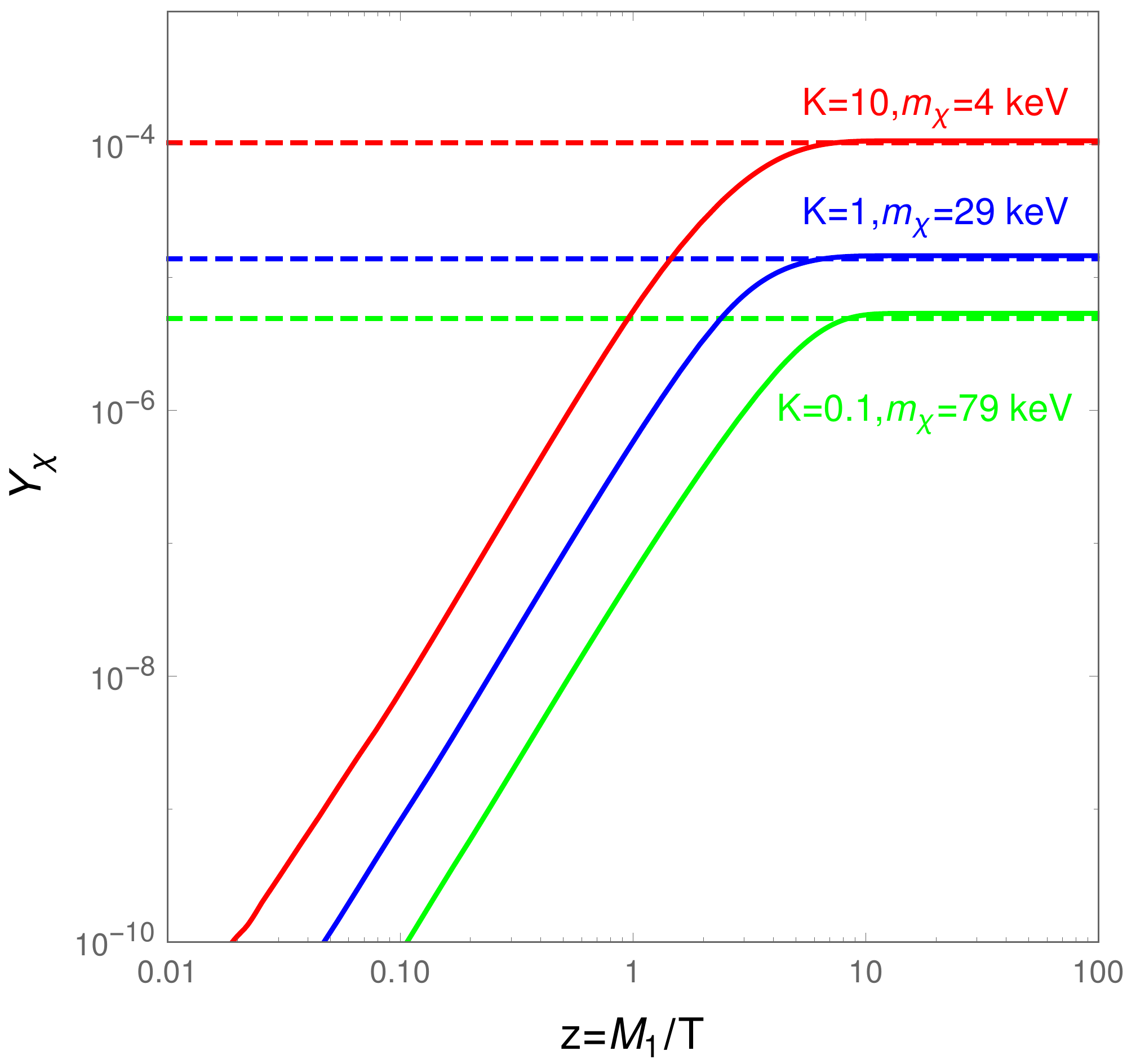}
\caption{ Evolution of dark matter abundance with parameter $z=M_1/T$. We fix $K=10$ in the left panel and BR$_\chi=10^{-3}$ in the right panel. The dashed horizontal lines correspond to the estimated results with Eq.~\eqref{eq:yx}. DM mass $m_\chi$ is obtained by setting $\Omega_\chi h^2=0.12$ with the numerical results of $Y_\chi(\infty)$.}
\label{FIG:YX-Z}
\end{figure}
According to the above Boltzmann equation, we can estimate the asymptotic abundances of $\chi$ as \cite{Falkowski:2017uya}
\begin{equation}\label{eq:yx}
Y_{\chi}(\infty)\simeq Y_{N_1}(0)\text{BR}_{\chi}\left(1
+\frac{15\pi\zeta(5)}{16\zeta(3)}K\right).
\end{equation}
Then, the corresponding relic abundance is
\begin{equation}\label{eq:ra}
\Omega_{\chi}{h^2}=\frac{m_\chi s_0 Y_\chi(\infty)}{\rho_c}h^2\simeq 0.12\times\left(\frac{m_{\chi}}{\keV}\right)
\left(\frac{\text{BR}_{\chi}}{10^{-3}}\right)\left(0.009+\frac{K}{44}\right),
\end{equation}
where $s_0=2891.2~\text{cm}^{-3}$, $\rho_c=1.05371\times10^{-5}h^2 ~\GeV~\text{cm}^{-3}$ \cite{Tanabashi:2018oca}. Typically, the observed relic abundance can be obtained with $m_\chi\sim4$ keV, $\text{BR}_\chi\sim10^{-3}$ and $K\sim10$. The evolution of DM abundances are shown in Fig.~(\ref{FIG:YX-Z}). It is clear that when the temperature goes down to $z=m_\chi/T\sim5$, the abundances $Y_\chi$ freeze in and keep at a constant. The left panel of Fig.~(\ref{FIG:YX-Z}) indicates that $m_\chi$ is inverse proportional to $\text{BR}_\chi$ when the decay parameter $K$ is a constant. For instance, sub-MeV scale light DM is obtained when $\text{BR}_\chi>10^{-6}$ with $K=10$.
Right panel of Fig.~(\ref{FIG:YX-Z}) shows the impact of decay parameter $K$.
Affected by the constant term before $K$ in Eq.~\eqref{eq:ra}, we can only conclude that the  smaller the $K$ is, the larger the $m_\chi$ is. Besides, we also find that the discrepancy between the numerical and analytical results of $Y_\chi(\infty)$ increases when $K$ decreases. Therefore, we adopt the numerical result of $Y_\chi(\infty)$ for a more precise calculation in the following discussion.

\begin{figure}
\centering
\includegraphics[width=0.45\textwidth]{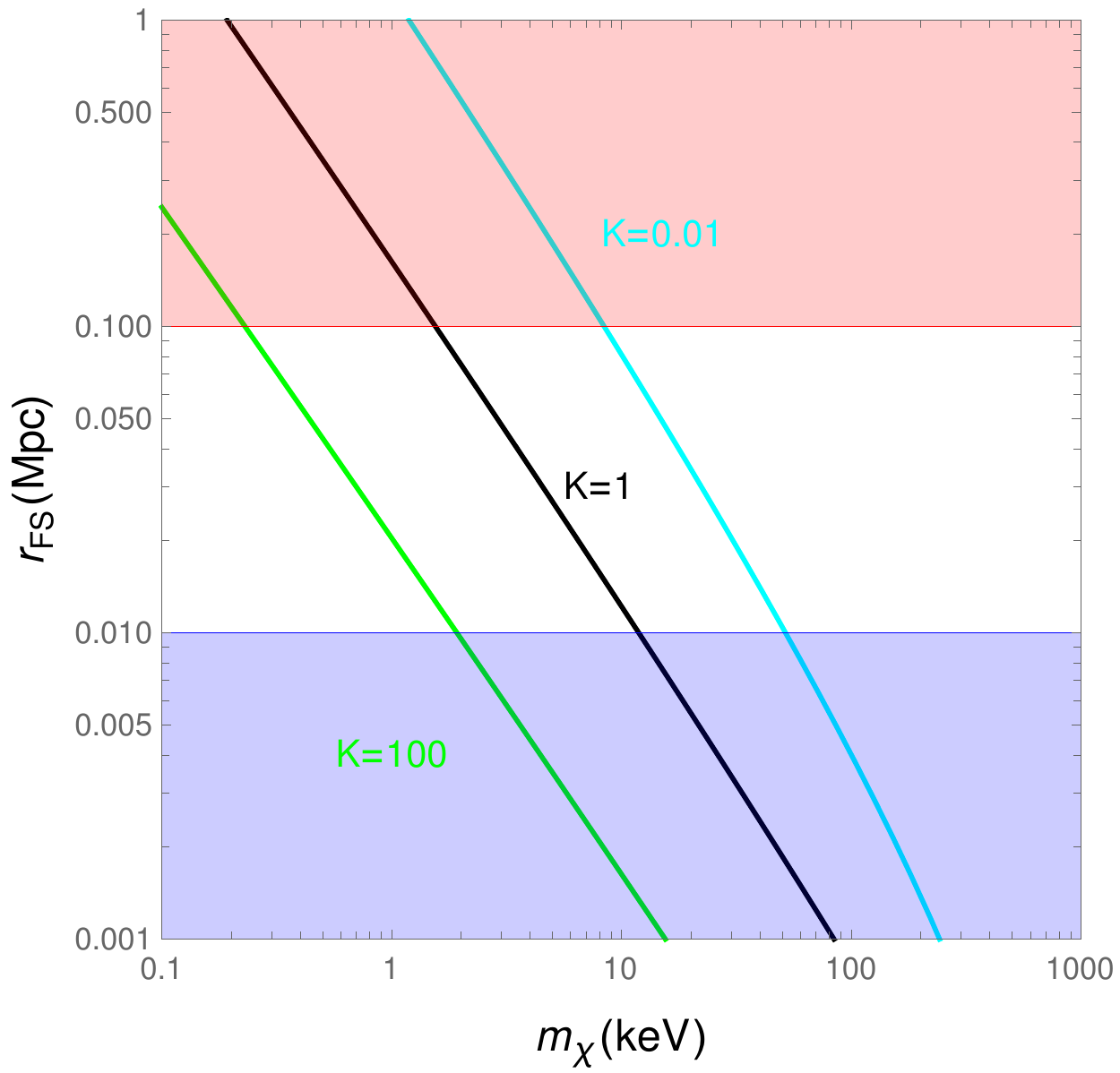}
\caption{Influence of free streaming on DM mass.  The red area ($r_{FS}>0.1$ Mpc), white area ($0.1~\text{Mpc}>r_{FS}>0.01$ Mpc) and blue area ($r_{FS}<0.01$ Mpc) correspond to hot, warm and cold DM scenario \cite{Merle:2013wta}, respectively.}
\label{FIG:rFS}
\end{figure}

The dominant constraint on FIMP DM $\chi$ comes from its free streaming length, which describes the average distance a particle travels without a collision \cite{Falkowski:2017uya}
\begin{equation}
r_{FS}=\int_{a_{rh}}^{a_{eq}}\frac{\langle v\rangle}{a^2H}da
\approx \frac{a_{NR}}{H_0\sqrt{\Omega_R}}
\left(0.62+\ln\left(\frac{a_{eq}}{a_{NR}}\right)\right),
\end{equation}
where $\langle v\rangle$ is the averaged velocity of DM $\chi$, $a_{eq}$ and ${a_{rh}}$ represent scale factors in equilibrium and reheating, respectively.
We use the results $H_0=67.3~\text{km}~  \text{s}^{-1}\text{Mpc}^{-1},\Omega_R=9.3\times10^{-5}$ and $a_{eq}=2.9\times10^{-4}$ obtained from Ref.~\cite{Ade:2015xua}.
The non-relativistic scale factor for FIMP DM is
\begin{equation}
a_{NR}=\frac{T_0}{2m_{\chi}}\left(\frac{g_{*,0}}{g_{*,rh}}\right)^{\frac{1}{3}}
K^{-\frac{1}{2}}.
\end{equation}
Taking $g_{*,0}=3.91$, $g_{*,rh}=106.75$ and $T_0=2.35\times 10^{-4}$ eV, finally we can get
\begin{equation}
r_{FS} \simeq2.8\times 10^{-2}\left(\frac{\keV}{m_{\chi}}\right)
\left(\frac{50}{K}\right)^{\frac{1}{2}}\times
\left(1+0.09\ln\left[\left(\frac{m_{\chi}}{\keV}\right)
\left(\frac{K}{50}\right)^{\frac{1}{2}}\right]\right) \text{Mpc}.
\end{equation}
The most stringent bound on $r_{FS} $ comes from small structure formation $r_{FS}<0.1$ Mpc \cite{Berlin:2017ftj}. The relationship between the mass of $\chi$ and its free streaming length is depicted in Fig.~(\ref{FIG:rFS}).
Basically speaking, warm DM is obtained for $m_\chi\sim10$ keV while $K\in[0.01,100]$. Meanwhile, $\chi$ becomes cold DM when $\chi$ is sufficient heavy and/or the decay parameter $K$ is large enough.

\section{Combined Analysis}\label{SEC:CA}

\begin{figure}
\begin{center}
\includegraphics[width=0.45\linewidth]{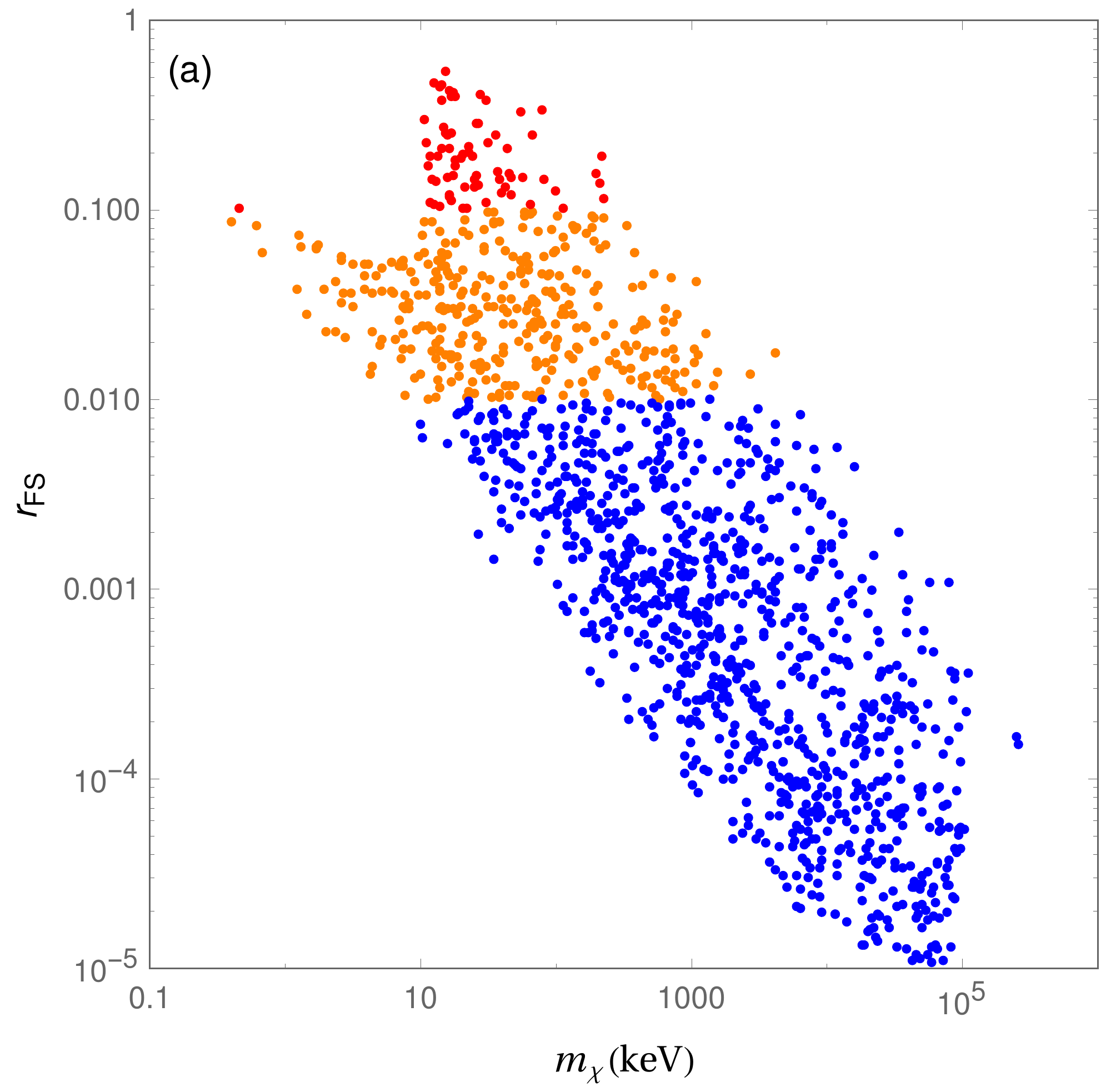}
\includegraphics[width=0.45\linewidth]{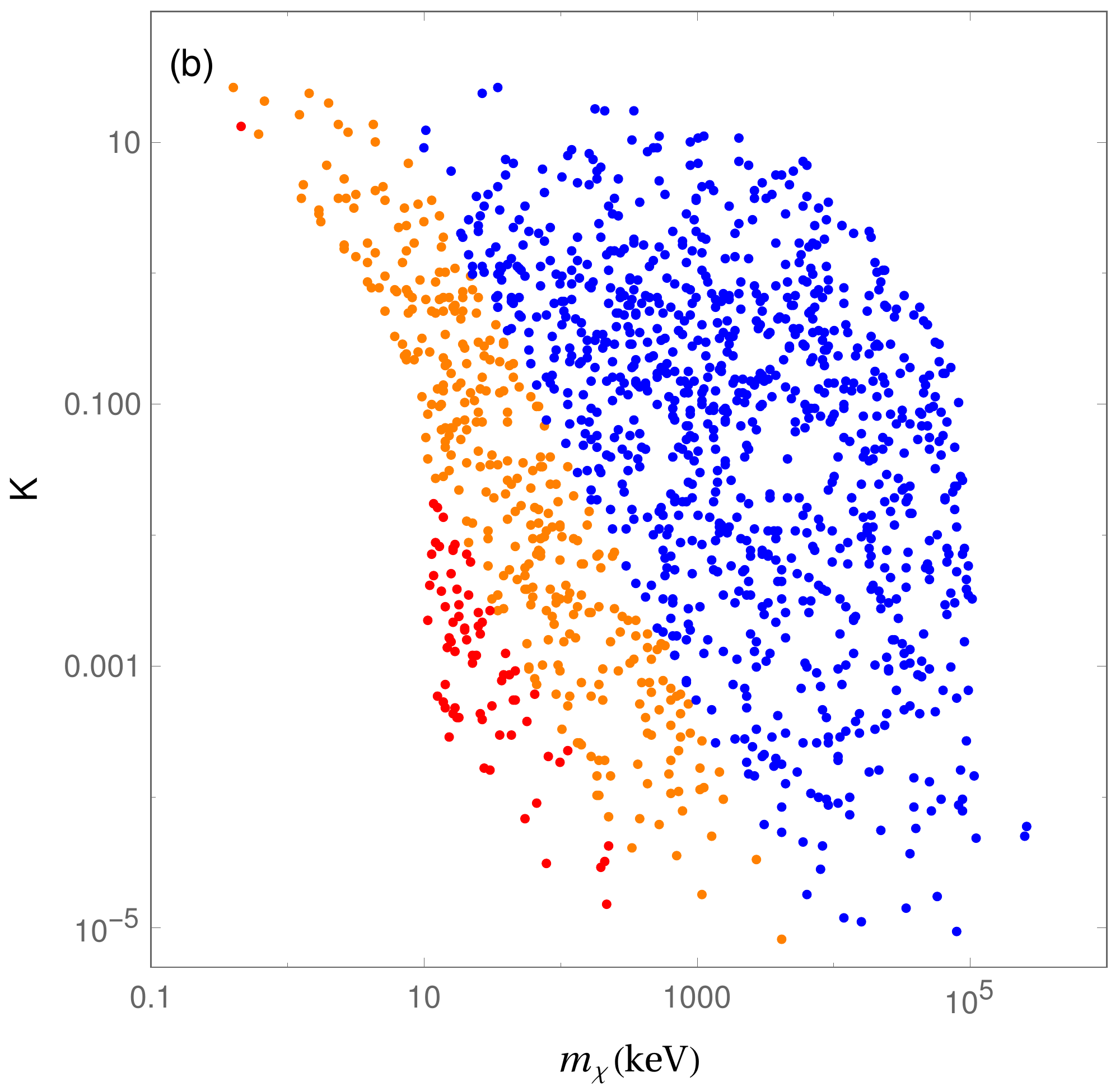}
\\
\includegraphics[width=0.45\linewidth]{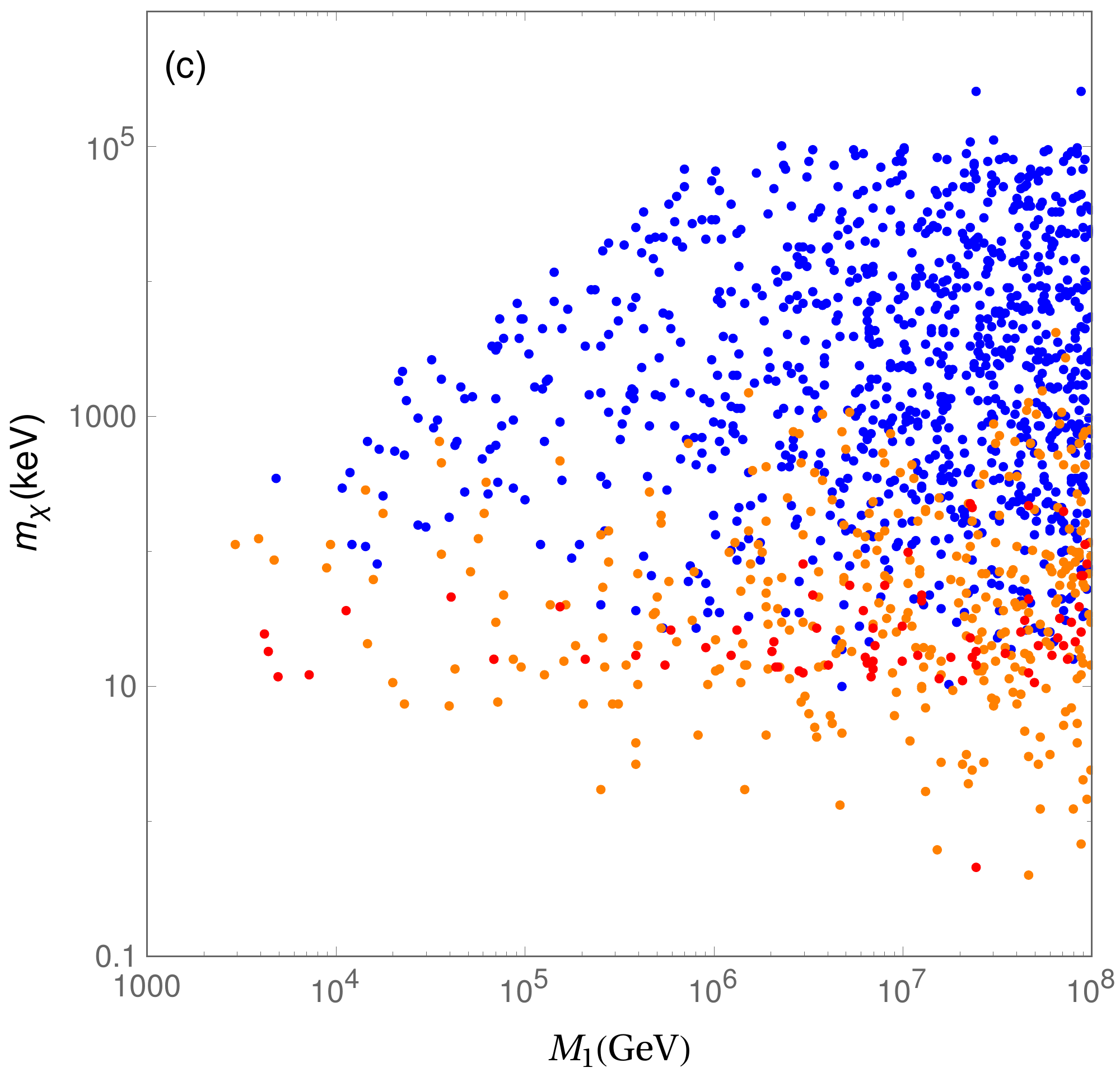}
\includegraphics[width=0.45\linewidth]{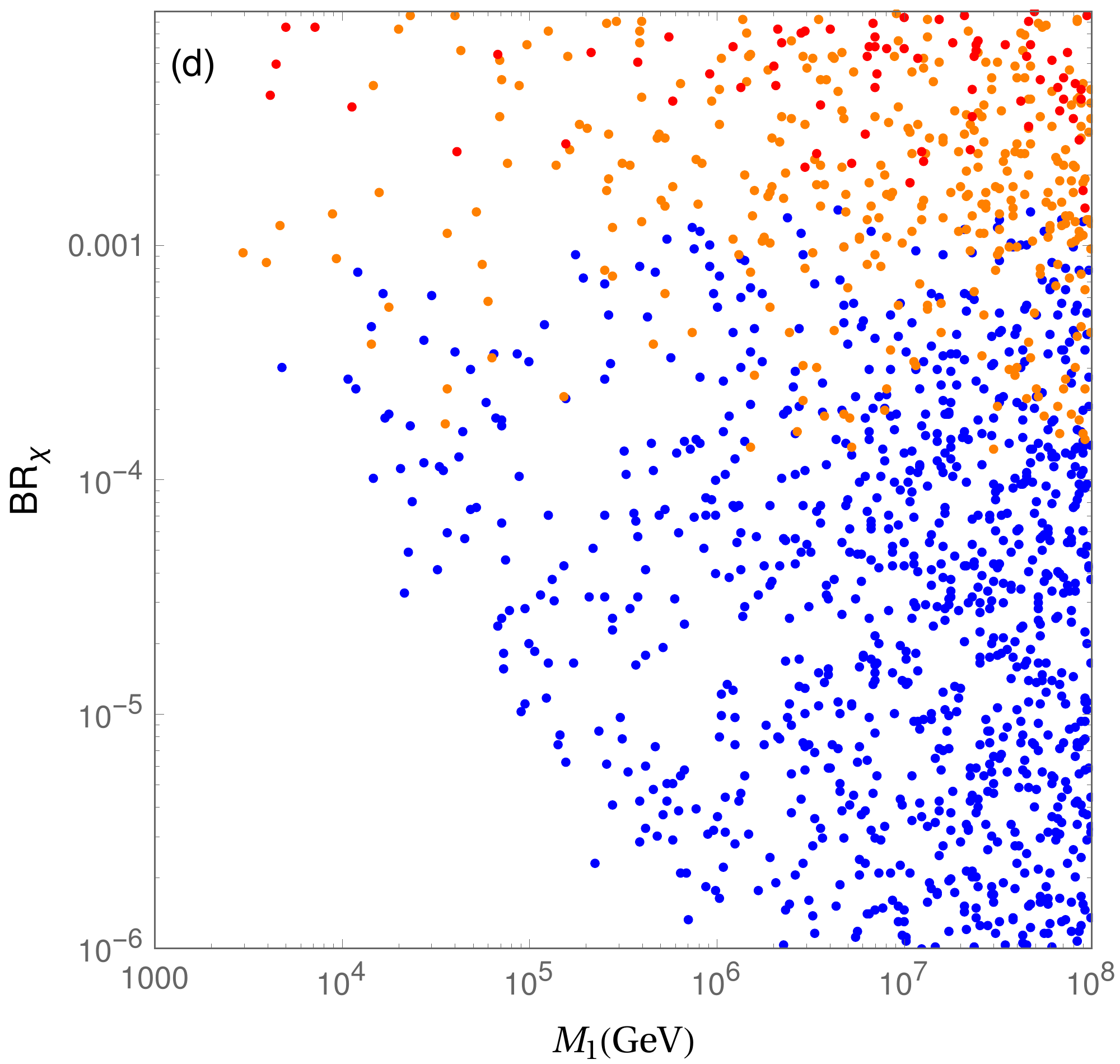}
\end{center}
\caption{Viable parameter space for DM. The red, orange and blue points correspond to hot, warm and cold DM, respectively.}
\label{FIG:DM}
\end{figure}

\begin{figure}
\begin{center}
\includegraphics[width=0.45\linewidth]{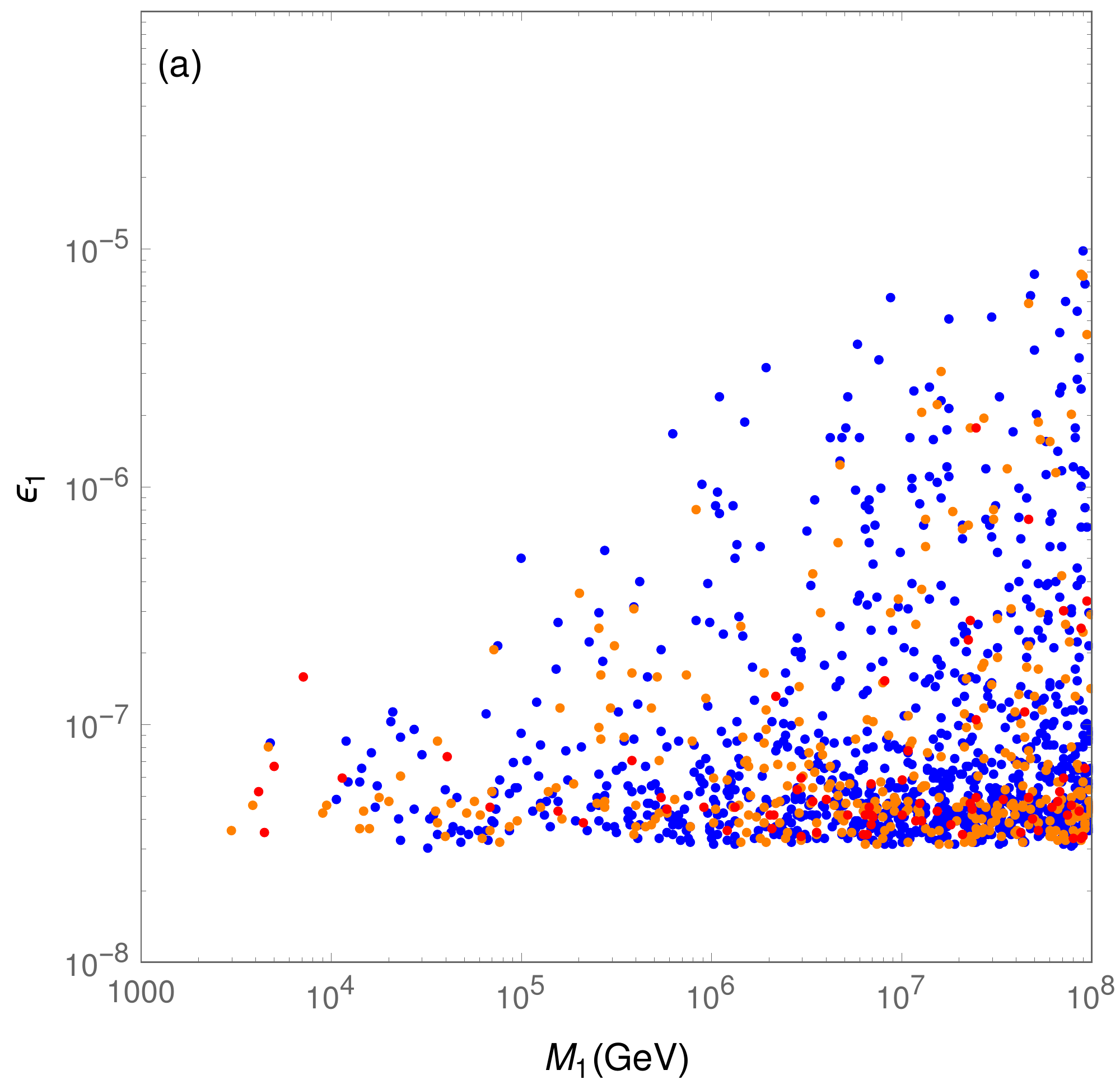}
\includegraphics[width=0.45\linewidth]{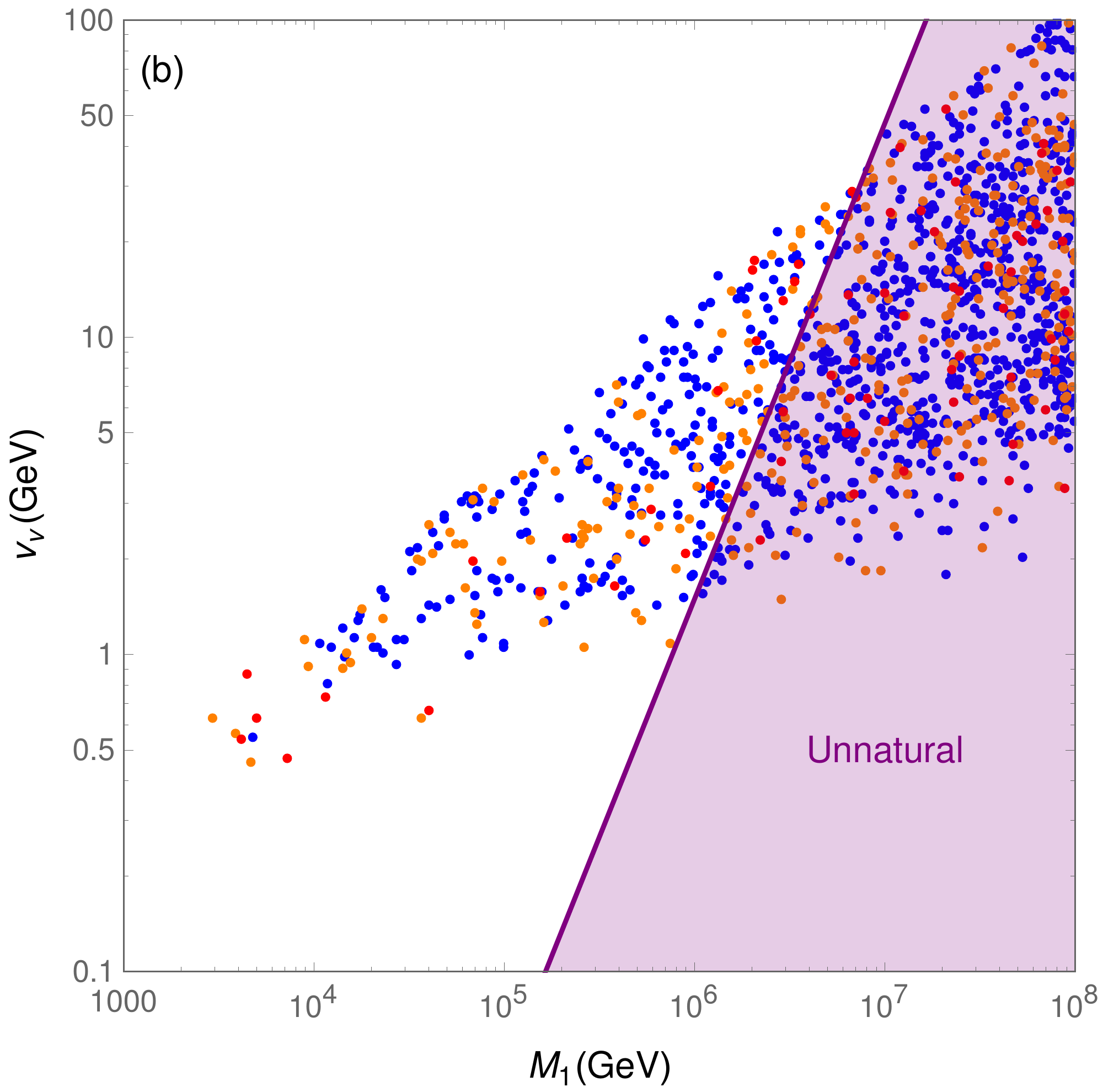}
\\
\includegraphics[width=0.45\linewidth]{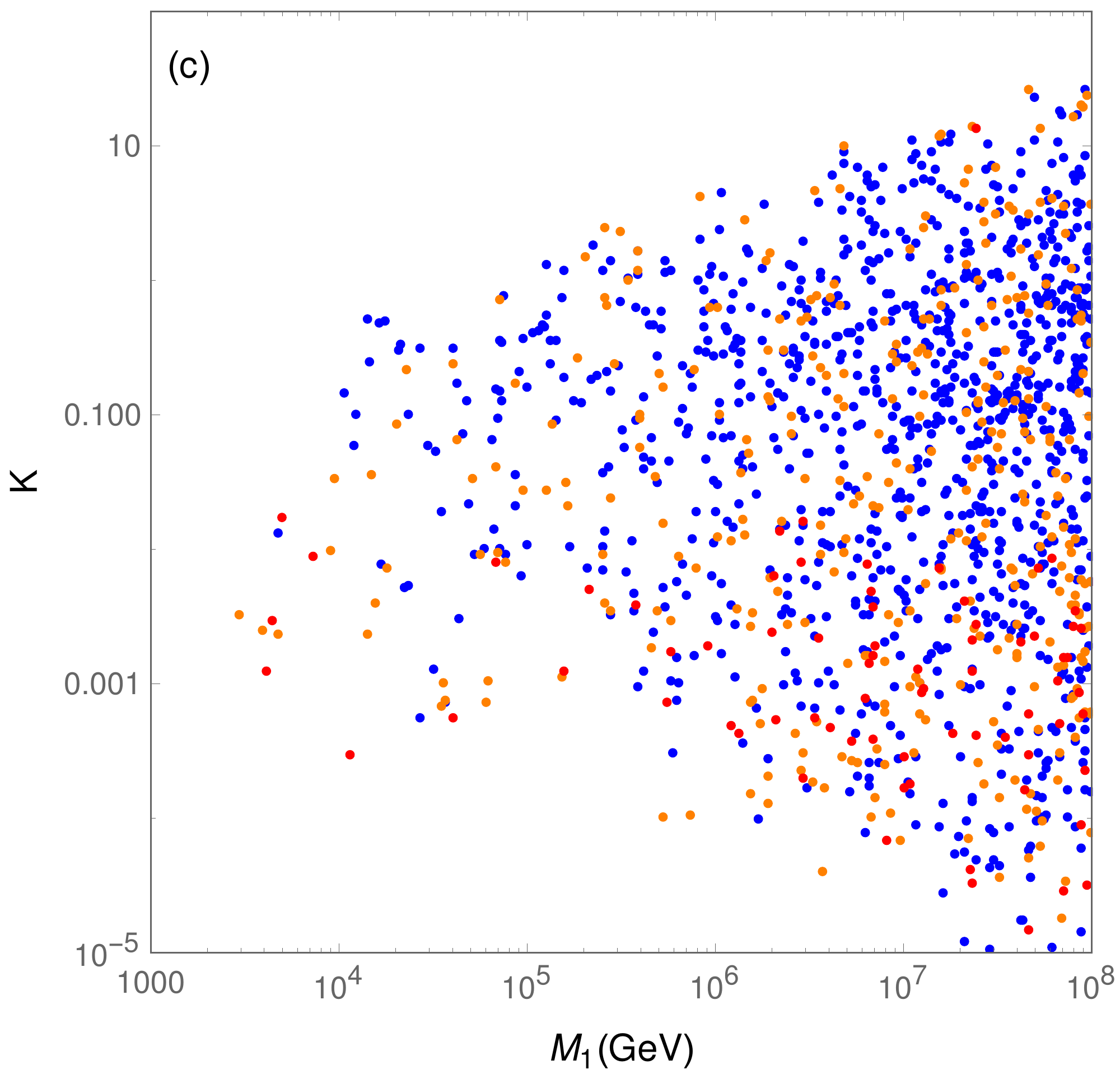}
\includegraphics[width=0.45\linewidth]{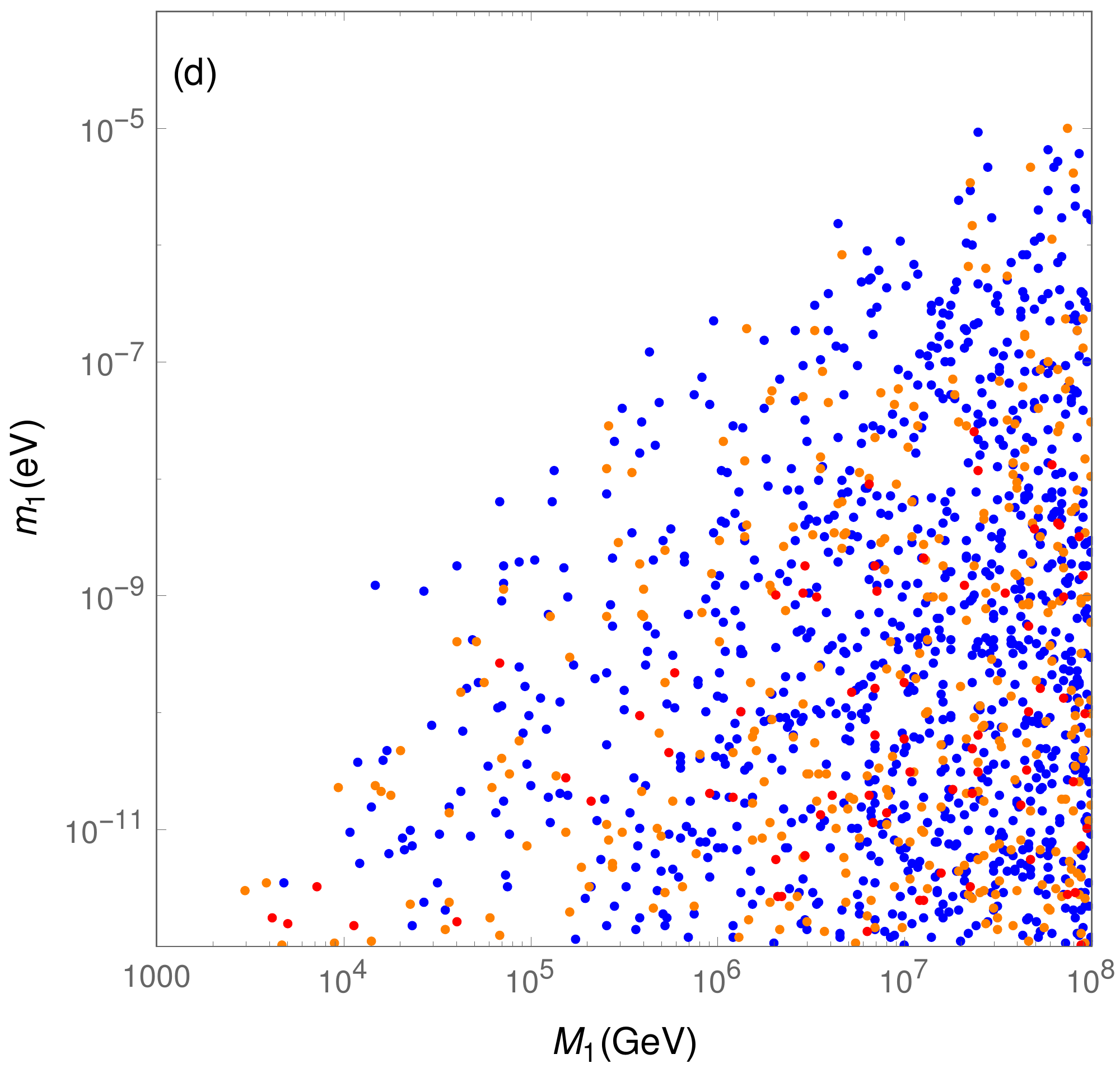}
\end{center}
\caption{Same as Fig~\ref{FIG:DM}, but for leptogenesis.}
\label{FIG:LG}
\end{figure}

After studying some benchmark points, it would be better to figure out the viable parameter space for success leptogenesis and DM. We then perform a random scan over the following parameter space:
\begin{eqnarray}
m_1\in[10^{-12},10^{-2}]~\eV, ~M_1\in[10^3,10^8]~\GeV, v_\nu\in[10^{-2},10^2]~\GeV,\\\nonumber
\text{Re}(\omega_{12,13,23})\in[10^{-10},1],~\text{Im}(\omega_{12,13,23})\in[10^{-10},1],~
\text{BR}_\chi\in[10^{-6},10^{-2}].
\end{eqnarray}
During the scan, we have fixed $M_2/M_1=M_3/M_2=10$. The final obtained baryon asymmetry $Y_{\Delta B}$ is required to be within $3\sigma$ range of the observed value, i.e., $Y_{\Delta B}\in[8.60,8.84]\times10^{-11}$. The results are shown in Fig.~\ref{FIG:DM} and Fig.~\ref{FIG:LG} for DM and leptogenesis, respectively.

Let's consider the DM results in Fig.~\ref{FIG:DM} first. According to the dominant constraint from  free streaming length $r_{FS}$, we can divide the viable samples into three scenarios in Fig.~\ref{FIG:DM} (a). Of course, the hot DM scenario is not favored by small structure formation. For warm DM, $m_\chi\in[0.3,2\times10^3]$ keV is possible. Meanwhile for cold DM, $m_\chi\in[10,2\times10^5]$ keV is allowed. And $r_{FS}$ is down to about $10^{-5}$ Mpc when $m_\chi\sim10^5$ keV. From Fig.~\ref{FIG:DM} (b), we aware that the hot DM samples correspond to those with small DM mass $m_\chi$  and very weak washout effect $K\lesssim10^{-2}$. Fig.~\ref{FIG:DM} (c) shows the samples in the $m_\chi-M_1$ plane. Three kinds of DM are all possible for certain value of $M_1$. By the way, it is interesting to obtain an upper limit on $m_\chi$ when $M_1\lesssim10^6$ GeV. This indicates that for TeV scale leptogenesis, FIMP DM should be keV to sub-MeV. The result for $\text{BR}_\chi$ is shown in Fig.~\ref{FIG:DM} (d), which tells us that warm DM requires $\text{BR}_\chi\gtrsim10^{-4}$ and cold DM requires $\text{BR}_\chi\lesssim10^{-3}$, respectively.

Then we consider the leptogenesis results in Fig.~\ref{FIG:LG}. The generalised  Davidson-Ibarra bound is clearly seen in Fig.~\ref{FIG:LG} (a). The (warm and cold DM) allowed samples show that the mass of $N_1$ for success leptogenesis could be down to about 3~TeV. The viable region in the $v_\nu-M_1$ plane is shown in Fig.~\ref{FIG:LG}~(b), which is consistent with the theoretical bounds discussed in Ref.~\cite{Clarke:2015hta}. For completeness, the naturalness bound in Eq.~\eqref{eq:M1} is also shown. Therefore, natural leptogenesis is viable for $3\times10^3~\GeV\lesssim M_1\lesssim7\times10^6$ GeV with $0.4~\GeV\lesssim v_\nu\lesssim30~\GeV$. The result for decay parameter $K$ is given in Fig.~\ref{FIG:LG}~(c), which shows that $K\lesssim10$ should be satisfied when $M_1\lesssim10^8$ GeV. Actually for $M_1\lesssim10^5$ GeV, all the samples are within weak washout region. An upper bound on lightest neutrino mass $m_1$ is clearly seen in Fig.~\ref{FIG:LG}~(d). Success leptogenesis in the $\nu$2HDM requires $m_1$ must be extremely tiny, i.e., $m_1\lesssim10^{-11}$ eV for $M_1\sim10^4$ GeV.

Before ending this section, we give a brief discussion on the collider signature.
According to the results of leptogenesis in Fig.~\ref{FIG:LG}, not too small $v_\nu$ is favored. In such scenario, the branching ratios of neutrinophilic scalars are quite different from the scenario with small $v_\nu$ \cite{Guo:2017ybk,Haba:2011nb,Wang:2016vfj,Huitu:2017vye}, but are similar with type-I 2HDM \cite{Branco:2011iw}. Currently, if $m_{\Phi_\nu}$ is smaller than $m_t$, the most stringent constraint comes from $t\to b H^\pm (H^\pm\to \tau^\pm\nu)$
\cite{Sirunyan:2019hkq}, which could exclude the region $v_\nu\gtrsim18$ GeV \cite{Sanyal:2019xcp}. Meanwhile, if $m_Z+m_h\lesssim m_{\phi_\nu}\lesssim2m_t$, the channel $A\to Zh(h\to b\bar{b})$ could exclude the region $v_\nu\gtrsim 24$ GeV \cite{Aaboud:2017cxo}. For heavier additional scalars with $m_{\Phi_\nu}>2 m_t$, the signature $A/H\to t\bar{t}$ is only able to probe the region $v_\nu\gtrsim 174$ GeV \cite{Aaboud:2017hnm,Chen:2019pkq}. Therefore, the experimental bounds on neutrinophilic scalars can be easily escaped provided $m_{\Phi_\nu}$ is large enough. At HL-LHC, the signature $A\to Zh(h\to b\bar{b})$ would reach $v_\nu\sim10$ GeV \cite{Chen:2019pkq}. Then the observation of this signature will indicate $M_1\sim10^6$ GeV and $m_1\lesssim10^{-7}$ eV.

\section{Conclusion}\label{SEC:CL}

In this paper, we propose an extended $\nu$2HDM to interpret the neutrino mass, leptogenesis and dark matter simultaneously. This model contains one neutrinophilic scalar doublet $\Phi_\nu$, three right hand heavy neutrino $N$,
which account for low scale neutrino mass generation similar to type-I seesaw.
Leptogenesis is generated due to the CP-violating decays of right hand neutrino $N\to \ell_L \Phi_\nu^*,\bar{\ell}_L \Phi_\nu$. The dark sector contains one scalar singlet $\phi$ and one Dirac fermion singlet $\chi$, which are charged under a $Z_2$ symmetry. Provided $m_\chi<m_\phi$ and $\lambda\ll1$, $\chi$ is a FIMP DM candidate within this paper. The relic abundance of $\chi$ is produced by $N\to \chi \phi$. Therefore, we have a common origin, i.e., the heavy right hand neutrino $N$, for tiny neutrino mass, baryon asymmetry and dark matter.

In the frame work of $\nu$2HDM, the asymmetry $\epsilon_1$ and  decay parameter $K$ are both enhanced by the smallness of $v_\nu$. By explicit calculation, we show that the decay parameter $K$ can be suppressed under certain circumstance.
The importance of $\Delta L=2$ washout process is also illustrated. As for FIMP DM, the relic abundance mainly depends on the branching ratio $\text{BR}_\chi$ and decay parameter $K$, and $m_\chi$ is typically at the order of keV to MeV scale. Meanwhile the free streaming length sets stringent bound. The viable parameter space for success leptogenesis and DM is obtained by solving the corresponding Boltzmann equations. To keep this model natural, we find $10^3~\GeV \lesssim M_1\lesssim10^6$ GeV, $0.4~\GeV\lesssim v_\nu\lesssim30~\GeV$, $m_1\lesssim10^{-5}$ eV and $K\lesssim10$ is favored by leptogenesis. Meanwhile, the warm (cold) DM mass in the range $m_\chi\in[0.3,2\times10^3]$ keV ($m_\chi\in[10,2\times10^5]$ keV) is predicted with $\text{BR}_\chi\gtrsim10^{-4}$ ($\text{BR}_\chi\lesssim10^{-3}$).

\section*{Acknowledgements}

This work is supported by National Natural Science Foundation of China under Grant No. 11805081, Natural Science Foundation of Shandong Province under Grant No. ZR2019QA021 and ZR2018MA047.



\begin{thebibliography}{000}

\bibitem{Fukuda:1998mi}
  Y.~Fukuda {\it et al.} [Super-Kamiokande Collaboration],
  Phys.\ Rev.\ Lett.\  {\bf 81}, 1562 (1998)
  [hep-ex/9807003].

\bibitem{Ahmad:2002jz}
  Q.~R.~Ahmad {\it et al.} [SNO Collaboration],
  Phys.\ Rev.\ Lett.\  {\bf 89}, 011301 (2002)
  [nucl-ex/0204008].

\bibitem{Minkowski:1977sc}
  P.~Minkowski,
  Phys.\ Lett.\ B {\bf 67}, 421 (1977).

\bibitem{Mohapatra:1979ia}
  R.~N.~Mohapatra and G.~Senjanovic,
  Phys.\ Rev.\ Lett.\  {\bf 44}, 912 (1980).

\bibitem{Fukugita:1986hr}
  M.~Fukugita and T.~Yanagida,
  Phys.\ Lett.\ B {\bf 174}, 45 (1986).

\bibitem{Davidson:2002qv}
  S.~Davidson and A.~Ibarra,
  Phys.\ Lett.\ B {\bf 535}, 25 (2002)
  [hep-ph/0202239].

\bibitem{Vissani:1997ys}
  F.~Vissani,
  Phys.\ Rev.\ D {\bf 57}, 7027 (1998)
  [hep-ph/9709409].

\bibitem{Clarke:2015gwa}
  J.~D.~Clarke, R.~Foot and R.~R.~Volkas,
  Phys.\ Rev.\ D {\bf 91}, no. 7, 073009 (2015)
  [arXiv:1502.01352 [hep-ph]].

\bibitem{Pilaftsis:2003gt}
  A.~Pilaftsis and T.~E.~J.~Underwood,
  Nucl.\ Phys.\ B {\bf 692}, 303 (2004)
  [hep-ph/0309342].

\bibitem{Akhmedov:1998qx}
  E.~K.~Akhmedov, V.~A.~Rubakov and A.~Y.~Smirnov,
  Phys.\ Rev.\ Lett.\  {\bf 81}, 1359 (1998)
  [hep-ph/9803255].

\bibitem{Asaka:2005pn}
  T.~Asaka and M.~Shaposhnikov,
  Phys.\ Lett.\ B {\bf 620}, 17 (2005)
  [hep-ph/0505013].

\bibitem{Hambye:2016sby}
  T.~Hambye and D.~Teresi,
  Phys.\ Rev.\ Lett.\  {\bf 117}, no. 9, 091801 (2016)
  [arXiv:1606.00017 [hep-ph]].

\bibitem{Hambye:2017elz}
  T.~Hambye and D.~Teresi,
  Phys.\ Rev.\ D {\bf 96}, no. 1, 015031 (2017)
  [arXiv:1705.00016 [hep-ph]].
  
\bibitem{Baumholzer:2018sfb}
  S.~Baumholzer, V.~Brdar and P.~Schwaller,
  JHEP {\bf 1808}, 067 (2018)
  [arXiv:1806.06864 [hep-ph]].

\bibitem{Chao:2012pt}
  W.~Chao and M.~J.~Ramsey-Musolf,
  Phys.\ Rev.\ D {\bf 89}, no. 3, 033007 (2014)
  [arXiv:1212.5709 [hep-ph]].
  
\bibitem{Clarke:2015hta}
  J.~D.~Clarke, R.~Foot and R.~R.~Volkas,
  Phys.\ Rev.\ D {\bf 92}, no. 3, 033006 (2015)
  [arXiv:1505.05744 [hep-ph]].

\bibitem{Ma:2006fn}
  E.~Ma,
  Mod.\ Phys.\ Lett.\ A {\bf 21}, 1777 (2006)
  doi:10.1142/S0217732306021141
  [hep-ph/0605180].
  
\bibitem{Kashiwase:2012xd}
  S.~Kashiwase and D.~Suematsu,
  Phys.\ Rev.\ D {\bf 86}, 053001 (2012)
  [arXiv:1207.2594 [hep-ph]].
 
\bibitem{Kashiwase:2013uy}
  S.~Kashiwase and D.~Suematsu,
  Eur.\ Phys.\ J.\ C {\bf 73}, 2484 (2013)
  [arXiv:1301.2087 [hep-ph]].

\bibitem{Racker:2013lua}
  J.~Racker,
  JCAP {\bf 1403}, 025 (2014)
  [arXiv:1308.1840 [hep-ph]].
     
\bibitem{Hugle:2018qbw}
  T.~Hugle, M.~Platscher and K.~Schmitz,
  Phys.\ Rev.\ D {\bf 98}, no. 2, 023020 (2018)
  [arXiv:1804.09660 [hep-ph]].

\bibitem{Borah:2018uci}
  D.~Borah, A.~Dasgupta and S.~K.~Kang,
  arXiv:1806.04689 [hep-ph].

\bibitem{Borah:2018rca}
  D.~Borah, P.~S.~B.~Dev and A.~Kumar,
  Phys.\ Rev.\ D {\bf 99}, no. 5, 055012 (2019)
  [arXiv:1810.03645 [hep-ph]].
  
\bibitem{Mahanta:2019sfo}
  D.~Mahanta and D.~Borah,
  arXiv:1912.09726 [hep-ph].
    
\bibitem{Ma:2000cc}
  E.~Ma,
  Phys.\ Rev.\ Lett.\  {\bf 86}, 2502 (2001)
  [hep-ph/0011121].


\bibitem{Haba:2011ra}
  N.~Haba and O.~Seto,
  Prog.\ Theor.\ Phys.\  {\bf 125}, 1155 (2011)
  [arXiv:1102.2889 [hep-ph]].

\bibitem{Haba:2011yc}
  N.~Haba and O.~Seto,
  Phys.\ Rev.\ D {\bf 84}, 103524 (2011)
  [arXiv:1106.5354 [hep-ph]].
  
\bibitem{Dodelson:1993je}
  S.~Dodelson and L.~M.~Widrow,
  Phys.\ Rev.\ Lett.\  {\bf 72}, 17 (1994)
  [hep-ph/9303287].

\bibitem{Adhikari:2016bei}
  M.~Drewes {\it et al.},
  JCAP {\bf 1701}, 025 (2017)
  [arXiv:1602.04816 [hep-ph]].

\bibitem{Adulpravitchai:2015mna}
  A.~Adulpravitchai and M.~A.~Schmidt,
  JHEP {\bf 1512}, 023 (2015)
  [arXiv:1507.05694 [hep-ph]].

\bibitem{Han:2018pek}
  Z.~L.~Han, B.~Zhu, L.~Bian and R.~Ding,
  arXiv:1812.00637 [hep-ph].

\bibitem{Boyarsky:2018tvu}
  A.~Boyarsky, M.~Drewes, T.~Lasserre, S.~Mertens and O.~Ruchayskiy,
  Prog.\ Part.\ Nucl.\ Phys.\  {\bf 104}, 1 (2019)
  [arXiv:1807.07938 [hep-ph]].

\bibitem{Antusch:2011nz}
  S.~Antusch, P.~Di Bari, D.~A.~Jones and S.~F.~King,
  Phys.\ Rev.\ D {\bf 86}, 023516 (2012)
  [arXiv:1107.6002 [hep-ph]].


\bibitem{Mahanta:2019gfe}
  D.~Mahanta and D.~Borah,
  JCAP {\bf 1911}, no. 11, 021 (2019)
  [arXiv:1906.03577 [hep-ph]].

\bibitem{Chianese:2019epo}
  M.~Chianese, B.~Fu and S.~F.~King,
  arXiv:1910.12916 [hep-ph].

\bibitem{Aprile:2018dbl}
  E.~Aprile {\it et al.} [XENON Collaboration],
  Phys.\ Rev.\ Lett.\  {\bf 121}, no. 11, 111302 (2018)
  [arXiv:1805.12562 [astro-ph.CO]].

\bibitem{Ackermann:2015zua}
  M.~Ackermann {\it et al.} [Fermi-LAT Collaboration],
  Phys.\ Rev.\ Lett.\  {\bf 115}, no. 23, 231301 (2015)
  [arXiv:1503.02641 [astro-ph.HE]].

\bibitem{Bernal:2017kxu}
  N.~Bernal, M.~Heikinheimo, T.~Tenkanen, K.~Tuominen and V.~Vaskonen,
  Int.\ J.\ Mod.\ Phys.\ A {\bf 32}, no. 27, 1730023 (2017)
  [arXiv:1706.07442 [hep-ph]].

\bibitem{Davidson:2009ha}
  S.~M.~Davidson and H.~E.~Logan,
  Phys.\ Rev.\ D {\bf 80}, 095008 (2009)
  [arXiv:0906.3335 [hep-ph]].

\bibitem{Morozumi:2011zu}
  T.~Morozumi, H.~Takata and K.~Tamai,
  Phys.\ Rev.\ D {\bf 85}, no. 5, 055002 (2012)
  Erratum: [Phys.\ Rev.\ D {\bf 89}, no. 7, 079901 (2014)]
  [arXiv:1107.1026 [hep-ph]].

\bibitem{Haba:2011fn}
  N.~Haba and T.~Horita,
  Phys.\ Lett.\ B {\bf 705}, 98 (2011)
  [arXiv:1107.3203 [hep-ph]].

\bibitem{Guo:2017ybk}
  C.~Guo, S.~Y.~Guo, Z.~L.~Han, B.~Li and Y.~Liao,
  JHEP {\bf 1704}, 065 (2017)
  [arXiv:1701.02463 [hep-ph]].

\bibitem{Aad:2012tfa}
  G.~Aad {\it et al.} [ATLAS Collaboration],
  Phys.\ Lett.\ B {\bf 716}, 1 (2012)
  [arXiv:1207.7214 [hep-ex]].

\bibitem{Chatrchyan:2012xdj}
  S.~Chatrchyan {\it et al.} [CMS Collaboration],
  Phys.\ Lett.\ B {\bf 716}, 30 (2012)
  [arXiv:1207.7235 [hep-ex]].

\bibitem{Machado:2015sha}
  P.~A.~N.~Machado, Y.~F.~Perez, O.~Sumensari, Z.~Tabrizi and R.~Z.~Funchal,
  JHEP {\bf 1512}, 160 (2015)
  [arXiv:1507.07550 [hep-ph]].

\bibitem{Casas:2001sr}
  J.~A.~Casas and A.~Ibarra,
  Nucl.\ Phys.\ B {\bf 618}, 171 (2001)
  [hep-ph/0103065].

\bibitem{Ibarra:2003up}
  A.~Ibarra and G.~G.~Ross,
  Phys.\ Lett.\ B {\bf 591}, 285 (2004)
  [hep-ph/0312138].

\bibitem{Nardi:2006fx}
  E.~Nardi, Y.~Nir, E.~Roulet and J.~Racker,
  JHEP {\bf 0601}, 164 (2006)
  [hep-ph/0601084].

\bibitem{Davidson:2008bu}
  S.~Davidson, E.~Nardi and Y.~Nir,
  Phys.\ Rept.\  {\bf 466}, 105 (2008)
  [arXiv:0802.2962 [hep-ph]].

\bibitem{Aghanim:2018eyx}
  N.~Aghanim {\it et al.} [Planck Collaboration],
  arXiv:1807.06209 [astro-ph.CO].
  
\bibitem{Falkowski:2011xh}
  A.~Falkowski, J.~T.~Ruderman and T.~Volansky,
  JHEP {\bf 1105}, 106 (2011)
  [arXiv:1101.4936 [hep-ph]].

\bibitem{Falkowski:2017uya}
  A.~Falkowski, E.~Kuflik, N.~Levi and T.~Volansky,
  Phys.\ Rev.\ D {\bf 99}, no. 1, 015022 (2019)
  [arXiv:1712.07652 [hep-ph]].

\bibitem{Buchmuller:2004nz}
  W.~Buchmuller, P.~Di Bari and M.~Plumacher,
  Annals Phys.\  {\bf 315}, 305 (2005)
  [hep-ph/0401240].

\bibitem{Esteban:2018azc}
  I.~Esteban, M.~C.~Gonzalez-Garcia, A.~Hernandez-Cabezudo, M.~Maltoni and T.~Schwetz,
  JHEP {\bf 1901}, 106 (2019)
  [arXiv:1811.05487 [hep-ph]].

\bibitem{Harvey:1990qw}
  J.~A.~Harvey and M.~S.~Turner,
  Phys.\ Rev.\ D {\bf 42}, 3344 (1990).

\bibitem{Hall:2009bx}
  L.~J.~Hall, K.~Jedamzik, J.~March-Russell and S.~M.~West,
  JHEP {\bf 1003} (2010) 080
  [arXiv:0911.1120 [hep-ph]].

\bibitem{Tanabashi:2018oca}
  M.~Tanabashi {\it et al.} [Particle Data Group],
  Phys.\ Rev.\ D {\bf 98}, no. 3, 030001 (2018).

\bibitem{Ade:2015xua}
  P.~A.~R.~Ade {\it et al.} [Planck Collaboration],
  Astron.\ Astrophys.\  {\bf 594}, A13 (2016)
  [arXiv:1502.01589 [astro-ph.CO]].

\bibitem{Berlin:2017ftj}
  A.~Berlin and N.~Blinov,
  Phys.\ Rev.\ Lett.\  {\bf 120}, no. 2, 021801 (2018)
  [arXiv:1706.07046 [hep-ph]].

\bibitem{Merle:2013wta}
  A.~Merle, V.~Niro and D.~Schmidt,
  JCAP {\bf 1403}, 028 (2014)
  [arXiv:1306.3996 [hep-ph]].

\bibitem{Haba:2011nb}
  N.~Haba and K.~Tsumura,
  JHEP {\bf 1106}, 068 (2011)
  [arXiv:1105.1409 [hep-ph]].

\bibitem{Wang:2016vfj}
  W.~Wang and Z.~L.~Han,
  Phys.\ Rev.\ D {\bf 94}, no. 5, 053015 (2016)
  [arXiv:1605.00239 [hep-ph]].

\bibitem{Huitu:2017vye}
  K.~Huitu, T.~J.~Karkkainen, S.~Mondal and S.~K.~Rai,
  Phys.\ Rev.\ D {\bf 97}, no. 3, 035026 (2018)
  [arXiv:1712.00338 [hep-ph]].
    
\bibitem{Branco:2011iw}
  G.~C.~Branco, P.~M.~Ferreira, L.~Lavoura, M.~N.~Rebelo, M.~Sher and J.~P.~Silva,
  Phys.\ Rept.\  {\bf 516}, 1 (2012)
  [arXiv:1106.0034 [hep-ph]].

\bibitem{Sirunyan:2019hkq}
  A.~M.~Sirunyan {\it et al.} [CMS Collaboration],
  JHEP {\bf 1907}, 142 (2019)
  [arXiv:1903.04560 [hep-ex]].

\bibitem{Sanyal:2019xcp}
  P.~Sanyal,
  Eur.\ Phys.\ J.\ C {\bf 79}, no. 11, 913 (2019)
  [arXiv:1906.02520 [hep-ph]].

\bibitem{Aaboud:2017cxo}
  M.~Aaboud {\it et al.} [ATLAS Collaboration],
  JHEP {\bf 1803}, 174 (2018)
  Erratum: [JHEP {\bf 1811}, 051 (2018)]
  [arXiv:1712.06518 [hep-ex]].

\bibitem{Aaboud:2017hnm}
  M.~Aaboud {\it et al.} [ATLAS Collaboration],
  Phys.\ Rev.\ Lett.\  {\bf 119}, no. 19, 191803 (2017)
  [arXiv:1707.06025 [hep-ex]].

\bibitem{Chen:2019pkq}
  N.~Chen, T.~Han, S.~Li, S.~Su, W.~Su and Y.~Wu,
  arXiv:1912.01431 [hep-ph].

\end{thebibliography}
\end{document}